# Characteristics and Predictive Modeling of Short-term Impacts of Hurricanes on the US Employment

Gan Zhang[1*] and Wenjun Zhu[2]

[1] Department of Climate, Meteorology, and Atmospheric Sciences, University of Illinois at Urbana-Champaign, 1301 W Green Street, Urbana, IL 61801, United States

[2] Nanyang Business School, Nanyang Technological University, 50 Nanyang Avenue, Singapore 639798

*Corresponding Author: Gan Zhang (gzhang13@illinois.edu)

# Abstract

The physical and economic damages of hurricanes can acutely affect employment and the well-being of employees. However, a comprehensive understanding of these impacts remains elusive as many studies focused on narrow subsets of regions or hurricanes. Here we present an open-source dataset that serves interdisciplinary research on hurricane impacts on US employment. Compared to past domain-specific efforts, this dataset has greater spatial-temporal granularity and variable coverage. To demonstrate potential applications of this dataset, we focus on the short-term employment disruptions related to hurricanes during 1990–2020. The observed county-level employment changes in the initial month are small on average, though large employment losses (>30%) can occur after extreme storms. The overall small changes partly result from a compensation among different employment sectors, which may obscure large, concentrated employment losses after hurricanes. Additional econometric analyses concur on the post-storm employment losses in hospitality and leisure but disagree on employment changes in the other industries. The dataset also enables data-driven analyses that highlight vulnerabilities such as pronounced employment losses related to Puerto Rico and rainy hurricanes. Furthermore, predictive modeling of short-term employment changes shows promising performance for service-providing industries and high-impact storms. In the examined cases, the nonlinear Random Forests model greatly outperforms the multiple linear regression model. The nonlinear model also suggests that more severe hurricane hazards projected by physical models may cause more extreme losses in US service-providing employment. Finally, we share our dataset and analytical code to facilitate the study and modeling hurricane impacts in a changing climate.

## 1. Introduction

Hurricanes, also known as tropical cyclones (TCs[1]), are rapidly rotating storm systems that can cause extreme wind, heavy rainfall, and severe floods. These storms can cause extensive life losses, property damages, and economic disruptions. Over the recent five decades, improvements in storm forecasting helped reduce global TC-related life losses by about 90% (World Meteorological Organization 2021). However, the direct economic losses caused by TCs stubbornly increased and reached greater than $500 billion during 2010-2019, making TCs the costliest among all the natural disasters (World Meteorological Organization 2021). As coastal development activities continue (Wilson and Fischetti 2010) and the physical risk of intense storms increases in a warming climate (Knutson *et al* 2020), TC-related monetary damages (Smith and Katz 2013, Smith 2020, Hemmati *et al* 2022), economic impacts (Hsiang 2010a, Strobl 2011, 2012, Hsiang and Jina 2014), and financial risks (Hereid 2022, Jerch *et al* 2023) have been extensively investigated on regional and global scales. While studies on the long-term socioeconomic impacts of TCs examine increasingly more storms ($O[10^2 \sim 10^3]$) (e.g., Hsiang 2010; Hsiang and Jina 2014; Tran and Wilson, 2020), the existing studies of the short-term socioeconomic impacts of TCs, such as employment shocks, mostly analyze a small number of cases and regions ($O[10^0 \sim 10^1]$). We posit that advancing the understanding of acute, short-term impacts of TCs can complement the studies of long-term socioeconomic impacts of hurricanes. More importantly, it may directly serve individuals and businesses in vulnerable communities by helping improve government aid programs and financial services (Bartik *et al* 2020).

Many socioeconomic and health impacts of TCs can be triggered or compounded by employment losses. For example, severe infrastructure failures can disrupt business activities and suppress employment opportunities. Potential unemployment can contribute to or worsen the financial stress for individuals, which has significant implications for public health (Galea *et al* 2007, Rhodes *et al* 2010, Parks *et al* 2021). When coupled with degrading post-storm living conditions (e.g., prolonged power outages), adverse employment changes can drive out-migration (McIntosh 2008, Strobl 2011), which in turn can worsen the local economy. Meanwhile, rebuilding activities can boost employment in certain sectors such as the construction industry (Groen *et al* 2020, Tran and Wilson 2020), which may partially offset the adverse impacts on employment. The existing estimates of post-storm employment changes vary substantially across studies and are hard to compare due to differences in metrics and subjects. For example, Tran and Wilson (2020) considered multiple natural disasters (e.g., hurricanes, tornadoes, and floods) in the US and found that the average employment changes within the first year after a disaster are about 1% in the construction section and <0.25% in the total nonfarm employment. Groen et al (2020) focused on areas affected by Hurricanes Katrina and Rita (2005) and reported about 10% population loss related to out-migration and about 15% of employment increase in the construction sector within one year. Wu et al. (2017) examined typhoons that affected Guangdong Province, China and reported a 12.5% increase in overall employment within one year. While labor market dynamics and storm characteristics may explain some differences, the reported employment changes have an unusually wide range that warrants further investigation. Without better knowledge of social vulnerability and the impact scale, it is challenging to effectively mobilize aids or services to serve the most vulnerable in disaster aftermaths.

---

[1] TCs in the North Atlantic and the East Pacific are termed as hurricanes. In these regions, hurricanes are also used by scientists and forecasters to denote the strong storms with a maximum wind speed of >64 knots. This study uses the term TC when discussing global storms but do not limit the use of hurricanes to intense storms.

Much of the current understanding of short-term employment changes after TCs was derived from storms that affected the US, which has substantial exposure to hurricanes and publishes open-access, reliable socioeconomics data. Despite intense investigation after Hurricane Katrina (2005), existing storm-employment studies generally focused on in-depth case analysis and strong hurricanes that make landfall in the states bordering the Gulf of Mexico (GoM). Unlike studies on hurricane-related long-term economic growth (Strobl 2011, Hsiang and Jina 2014), the existing hurricane-employment studies are concentrated on a narrow subsets of several high-impact hurricanes (e.g., Hurricane Katrina) (Brown *et al* 2006, Groen and Polivka 2008, McIntosh 2008, Groen *et al* 2020) and limited states in the U.S. (e.g., Florida) (Belasen and Polachek 2008, Goulbourne 2021, Groen *et al* 2020). Consequently, the existing studies have not fully accounted for the regional heterogeneity of communities or the diversity of storm characteristics.

While a focus on the GoM states has afforded valuable insights, it leaves many hurricane-prone communities underserved. These include the East Coast states, inland communities, and unincorporated territories (e.g., Puerto Rico), many of which are vulnerable to hurricane hazards. For example, water-isolated and less resourceful regions like Puerto Rico might also be less resilient, as suggested by the shipping disruptions and extended blackout after Hurricane Maria (2017). Additionally, unique socioeconomic factors, such as weaker building codes or heavy dependence on specific economic industries (e.g., leisure and hospitality), may make employment in some understudied communities more sensitive to hurricane impacts. For example, Hsiang (2010) showed investigated the economic production in the Caribbean and Central America and found hurricanes have persistent, negative impacts on the production by tourism and agriculture. While it is hard for a single study to address all those topics, identifying regional vulnerabilities that received limited attention can be meaningful.

Potential sensitivities of employment changes to storm characteristics also warrant attention. Empirical evidence suggests the economic impacts of hurricanes scale nonlinearly with the storm intensity (Nordhaus 2010). When it comes to employment-related findings, whether the findings from a small group of intense storms apply to many weaker storms is unclear. Another potential limitation of past employment studies is their exclusiveness with the metric of maximum wind speed. At higher latitudes, hurricanes tend to experience structural changes that can broaden their wind field (e.g., Evans *et al* 2017) and cause heavy damages even at relatively low intensity (e.g., Hurricane Sandy 2012). For other storm hazards, such as heavy precipitation (e.g., Hurricane Harvey 2017), it is unclear whether a wind-based metric fully characterize their potential impacts. Hurricane-related extreme precipitation often affect inland regions past by received little attention by hurricane-employment studies. For inland regions, wind hazards associated with decaying hurricanes greatly weaken, but extreme rainfall can occur >1500 km away from the landfall point and cause heavy damages (e.g., Hurricane Ida 2021). Understanding whether and how such mid-latitude or inland incidents might affect employment may benefit large populations.

A comprehensive understanding of those diverse sensitivities is a prerequisite to addressing emerging challenges related to long-term changes in socioeconomics and climate. The coastal population has been increasing in the US (Wilson and Fischetti 2010), suggesting that employment and economic activities with hurricane exposure will likely increase. Meanwhile, the physical risk can also change with the climate. For example, a warmer climate will lead to sea level rise, more extreme rainfall, and a likely increase in TC intensity (Knutson *et al* 2020, Seneviratne *et al* 2021). Some physical evidence also suggests a poleward expansion of TC activity to the midlatitudes (e.g., Kossin *et al* 2014, Zhang *et al* 2021, Studholme *et al* 2022) and a slowdown in storm motion that might execrate localized hazard exposure (e.g., Kossin 2018; Zhang et al. 2020). These changes suggest more severe TC hazards might increasingly occur in diverse socioeconomic contexts in the future.

Here we focus on the US to leverage its high-quality and open-access data. This focus is also motivated by the fact that about 65% of cyclone-related economic damages occurred in North America (World Meteorological Organization 2021) and that hurricanes stand out as the leading cause of US billion-dollar disasters (Smith and Katz 2013). Besides providing an open-access dataset with physical and socioeconomic variables, this hurricane-employment study also strives to promote interdisciplinary research and serve more diverse communities. While our analytics do not fully address all the outstanding research questions reviewed earlier, we embrace the open science practice and hope our efforts may serve as an initial step which follow-up studies may build on. The rest of this manuscript is organized as follows. Section 2 provides an overview of the data and methodology. Section 3 presents descriptive analyses of the employment data with an emphasis on notable hurricane cases and various hazard conditions. Section 4 follows the practice of causality inference and analyzes the employment changes more rigorously. Section 5 adopts a data-driven perspective and highlights potential applications of this dataset. The study concludes with a summary and discussions.

## 2. Data and Methodology

### 2.1 Data

The hurricane data is from the International Best Track Archive for Climate Stewardship (IBTrACS; Knapp *et al* 2010). The dataset includes the time, location, maximum wind, and other physical characteristics of Atlantic storms, which are available at a 3-hourly resolution. Since IBTrACS does not include precipitation information, we use the daily precipitation data from the Global Unified Gauge-Based Analysis (Chen *et al* 2008) by the US Climate Prediciton Center (CPC). While higher-resolution precipitation data is available, they generally contain large uncertainties (e.g., reanalysis datasets) or have limited spatial-temporal coverage (e.g., satellite datasets). The Global unified gauge data were derived from in-situ station observations and interpolated to a grid of 0.5-degree spacing. The size of individual grid cells is comparable to the average size of the US counties.

The US Bureau of Labor Statistics (BLS) provides employment data. Specifically, we use the Quarterly Census of Employment and Wages (https://www.bls.gov/cew/downloadable-data-files.htm) which covers the government-defined industry sectors and includes monthly employment data and quarterly wage data. The dataset covers all the county-level entities in the US and individual metropolitan statistical areas (MSAs), which are aggregations of county-level entities with close socioeconomic ties. This study processed county-level and MSA-level data from 1990 to 2021, but we conduct most analyses on the county level as other socioeconomic data at the MSA level has limited availability. The original BLS employment dataset is well maintained but still contains minor errors (e.g., missing records), especially in regions that have a relatively small number of employees. Accordingly, we drop the entries with missing records or unrealistic values. To limit the possible impacts of data errors related to small employment bases, our analyses focus on entities with at least 100 employees in the goods-producing and service-proving sectors.

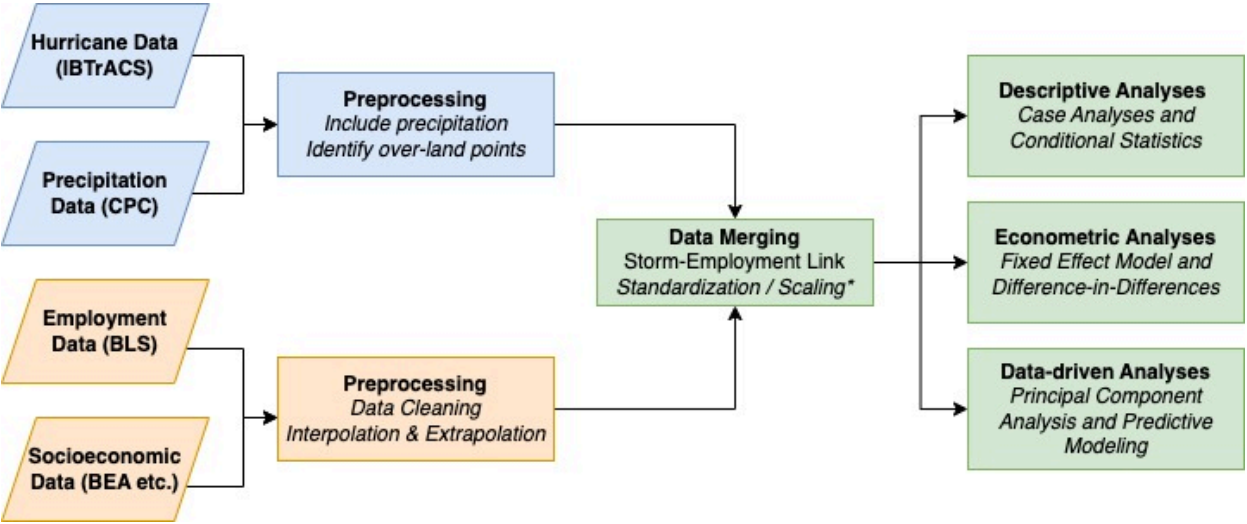

**Figure 1 Flow chart of data, preprocessing, and analyses. The datasets and the intermediate steps are introduced in Section 2. The descriptive analyses, econometric analyses, and data-driven analyses correspond to Sections 3, 4, and 5, respectively. Some of the analyses involve data standardization or scaling, which will be described separately in individual sections.**

Additional socioeconomic data including the population, education, and personal income data are available at the US county level. The links to the original data sources are provided in the Data Availability Statement. Specifically, we acquire the population data from the Division of Cancer Control and Population Sciences (DCCPS) affiliated with the US National Institute of Health. The original population data is from the US Census. This study uses the DCCPS version as it organizes and aggregates age information in more accessible formats that facilitate the evaluation of the working-age population (15–64 years old). The personal income data is from economic profile (CAINC30) of the U.S. Bureau of Economic Analysis (BEA). We considered the Gross Domestic

Product (GDP) per capita as an alternative, yet the GDP data is unavailable at the county level before 2000. The education level data is from the preprocessed data by US Department of Agriculture (USDA). We aggregate the percentage of the population that hold degrees at or above the high-school level. The population and income data are available at an annual base, while the education data is available at a quasi-decadal base. For the econometric analyses, we linearly interpolated the slow-varying socioeconomic data to match the monthly frequency of employment data. For the education data, the last available time window is 2017–2021, so we assign the average value to 2019 and linearly extrapolate the data to 2020 and 2021.

**2.2 Methodology**

This study analyzes the data described in Section 2.1 following the workflow outlined by Figure 1. The workflow prefers conceptually simple and computationally affordable methods, which helps deliver a "minimum viable product" that future research may iterate and build on. The preference deviates from some existing economics studies (details in Section 2.2.1) but enables us to analyze storm-employment data with great granularity and scale ($O[10^5]$ versus $O[10^{1-2}]$). We will next describe how individual storms are linked to employment changes, as well as analytical and modeling techniques. Additional details about the data processing are available in Supplementary Information and the open-access repository (Data Availability Statement).

*2.2.1 Linking Storms to Employment and Socioeconomic Data*

Recent economics studies have attempted to characterize hurricane impacts using wind fields derived from idealized physical models (e.g., Hsiang 2010, Deryugina 2017). However, such idealized models are not straightforward to apply to the precipitation field, which is governed by different physical equations. Additionally, the physical approximations of idealized wind models can be violated when hurricanes move over land or experience structural changes at the mid-latitudes, such as the extratropical transition (e.g., Evans *et al* 2017). Those issues are further complicated by challenges related to the data volume and computational burden.

Given those complications, we instead link individual storms (and their hazards) to employment changes at geographic entities (e.g., counties) by matching their spatial-temporal information. For each storm data point over land, we search for geographic entities with a geometric centroid that is within 200 km of the storm center. The radius threshold is based on the typical TC geometry and real-world experience in assessing storm damages. Increasing the distance threshold introduces more samples of little direct impacts, while reducing the threshold to <100 km misses the impacts of large storms (e.g., Hurricane Sandy 2012). When the distance threshold is satisfied, the entity is considered storm-affected, and the related physical and socioeconomics information is extracted. The information includes storm data, geographic metadata, labor market data, and so on. This workflow ignores the geospatial details of hazards related to individual storms, which we assume are of secondary importance and leave for future studies. Lastly, a close examination reveals a weakness of the over-land masking method with storms meandering offshore. Nonetheless, this issue affects only a small number of data points and is unlikely to undermine the overall analysis (Supplementary Information).

At the county level, the employment and socioeconomic data are not perfectly matched owing to differences in geographic aggregations by US agencies. For example, the aggregation of counties and independent cities differs between data sources in Virginia. Some socioeconomic datasets did not include Puerto Rico, a territory that is vulnerable to hurricanes but tends to be underserved. To minimize differences related to the geographic aggregation, we limit the use of non-employment socioeconomic data to econometric analyses only (Section 4). When record aggregation does not match, we drop all the related data records to avoid ambiguity.

*2.2.2 Analyses of Employment Changes*

This study adopts an open-access, interdisciplinary practice and emphasizes the accessibility and replicability of analyses. The analyses of employment changes consist of three complementary parts: descriptive analyses, econometric analyses, and data-driven analyses. The descriptive analyses follow previous studies of short-term employment changes and start with several cases of destructive hurricanes. The case selection prioritizes the representation of a relatively broad range of communities and storms. The case analysis is augmented by the statistical analysis of all the relevant incidents in 1990–2020, which help highlight the scope and extremes of post-hurricane employment changes. These analyses are followed by composite analysis that provides insights into sector-level employment sensitivity to storm hazards. The descriptive analyses serve to characterize the new dataset and motivate the ensuing analytics.

The econometric analyses extend to the causality inference and include a fixed effects model and the difference-in-differences (DID) analyses. The analyses consider an extensive set of socioeconomic variables (Section 2.1) and excluded Puerto Rico and some Virginia entities due to data availability (Section 2.2.1). Using these variables, we model the county-level employment using the following fixed effects model:

$$y_{it} = \alpha_i + \lambda_t + \delta D_{it} + \beta X_{it} + \epsilon_{it} \quad (1)$$

where $y_{it}$ is the $log_{10}$ of employment number, and the subscript $i$ and $t$ denote individual entities and time steps, respectively. The parameter $\alpha_i$ is unobserved time-invariant individual effect, and $\lambda_t$ is the location-independent time effect. The parameter $D_{it}$ is the dummy variable of hurricane passing (defined in Section 2.2.1), and $\delta$ is the hurricane-associated regression coefficient. The vector $X_{it}$ is the observed time-varying covariates (e.g., income per capita, work-age population, and education level), and $\beta$ is the associated regression coefficient vector. The $\epsilon_{it}$ is an error term of the fixed effects model, and the model fitting uses demeaned variables. This basic model helps evaluate potential hurricane impacts, and its results are compared to those from the differences-in-differences (DiD) analysis. The DiD analysis is conducted in the double / debiased machine learning framework (Chernozhukov *et al* 2018) using the recipe by the DoubleML package (Bach *et al* 2022). We evaluate the average treatment effect related to hurricanes under the assumption of conditional parallel trend assumption. The DiD evaluation is conducted for the period four months before and twelve-month after hurricane passing. For each month with a hurricane, we build snapshots of the county-level entities across the US and append all the available data records. To mitigate complications related to earlier or later hurricane hits, we only consider samples with hurricane passing at Month 0 or without any hurricane passing in the -4 to +12 month window.

The data-driven analytics provide examples of pattern identification and predictive modeling. We adopt the principal component analysis (PCA) to extract patterns and regression models to explore the possibility of predicting employment changes. As a standard pre-processing step, we follow the common practice of data science and standardize all the input variables (i.e., removing the mean and scaling to unit variance). PCA is a common technique of dimension reduction (Jolliffe and Cadima 2016) that can help explore covarying modes of storm features and employment changes. The PCA may yield modes that are aligned with the existing knowledge or novel modes that received little attention. We also explore the predictive modeling of employment shocks based on simple linear and nonlinear algorithms (i.e., Random Forests; Breiman 2001). The simple predictive models are validated using a five-fold cross-validation to demonstrate that hurricane-related employment shocks can be robustly predicted. We also consider possible hazard changes related to climate change. By applying a fractional change to historical hurricane observations, we use the Random Forrests model to infer possible post-storm employment changes in future climate scenarios.

## 3. Descriptive Analyses of Post-Hurricane Employment Changes

### 3.1 Employment Changes Associated with Extreme Hurricanes

To validate our methods and demonstrate post-storm employment changes, we first analyze six hurricanes that resulted in severe economic damages (Table 1). The case choice prioritizes the representation of diverse communities affected by high-damage, recent hurricanes. For example, the selected cases cover communities with various population sizes (e.g., megacities and medium-sized communities) and geographic characteristics (e.g., subtropics and midlatitude). We examine employment changes at the county and MSA levels. The hazards of the examined storms include extreme wind, heavy rainfall, and storm surges. When a hurricane makes multiple landfalls (e.g., Hurricanes Andrew and Harvey), the county-level analysis considers the location where the initial landfall occurs, and MSA analysis targets the nearest entities where significant economic damages occurred.

**Table 1 Characteristics of analyzed hurricanes and regions. The economic losses are inflation-adjusted and provided by NCEI (Smith 2020).**

| Cyclone Name | Economic Losses | Major Hazards | Region Profile |
|---|---|---|---|
| **Andrew (1992)** | **$59.4 Billion** | High wind | Large metropolitan, subtropical coast |
| **Katrina (2005)** | **$186.3 Billion** | High wind, storm surge, flood | Large metropolitan, subtropical coast |
| **Sandy (2012)** | **$81.9 Billion** | Midlatitude, storm surge, flood | Large metropolitan, mid-latitude coast |
| **Harvey (2017)** | **$148.8 Billion** | High wind, heavy rainfall, flood | Large metropolitan, subtropical coast |
| **Laura (2020)** | **$26.0 Billion** | High wind, storm surge, flood | Small metropolitan, subtropical coast |
| **Maria (2017)** | **$107.1 Billion** | High wind, heavy rainfall, flood | Large metropolitan, tropical island |

Figure 2 suggests that the total employment in the analyzed high-damage cases generally reduced after storm impacts. The immediate employment (Month 1) shock ranges from about -30% to near-zero. The most severe employment loss occurred in New Orleans-Metairie LA (Fig. 2h) after Katrina (2005), likely due to devastating infrastructure failures. Similar employment losses appeared at the county level in several other cases, including Hurricane Harvey (2017) and Hurricane Laura (2020). While non-storm factors may complicate the interpretation of post-storm employment changes, a comparison with the employment before hurricane hit suggests a hurricane-related shock can be evident even amidst the extreme economic changes (e.g., pandemic lockdown before Hurricane Laura) (Fig. 2f). Perhaps surprisingly, the storm-affected MSAs—despite apparent economic losses and infrastructure disruptions—show minor or even near-zero employment changes. In the cases where employment changes are large ($^3$20%), the time of the employment recovery ranges from about three months to more than one year. The exceptionally slow recovery from Hurricane Katrina (2005) is consistent with findings by previous studies (e.g., Garber et al. 2006). Such slow recovery of employment also occurred in other cases that received much less attention (e.g., Fig. 2d).

Figure 2 also suggests that the total changes in post-storm employment are mainly contributed by the private sector. Considering the labor and operation practices, private-sector employment is likely more responsive to acute disruptions. For example, Bartik *et al* (2020) reported that the median cash reserve of small businesses in the US can sustain operation for about two weeks, which makes these employers vulnerable to prolonged business disruptions. Recent business surveys suggest about 60% of small businesses close at least temporarily after disasters (Hiti *et al* 2022). The US Federal Emergency Management Agency (FEMA) suggested that 43% of small businesses affected by a disaster never reopen (FEMA n.d.). In comparison to the private-sector employment, the connection between the changes in the government employment and storm impacts are less clear. Nonetheless, the government sector loss is noteworthy after Hurricane

Katrina (2005), possibly due to a reduction in tax revenues and expenditures, and an increase the cost of borrowing after disasters (Jerch *et al* 2023). However, the availability of BLS data appears less consistent for government employment (Fig. 2). Accordingly, the ensuing analyses will focus on employment changes in the private sector.

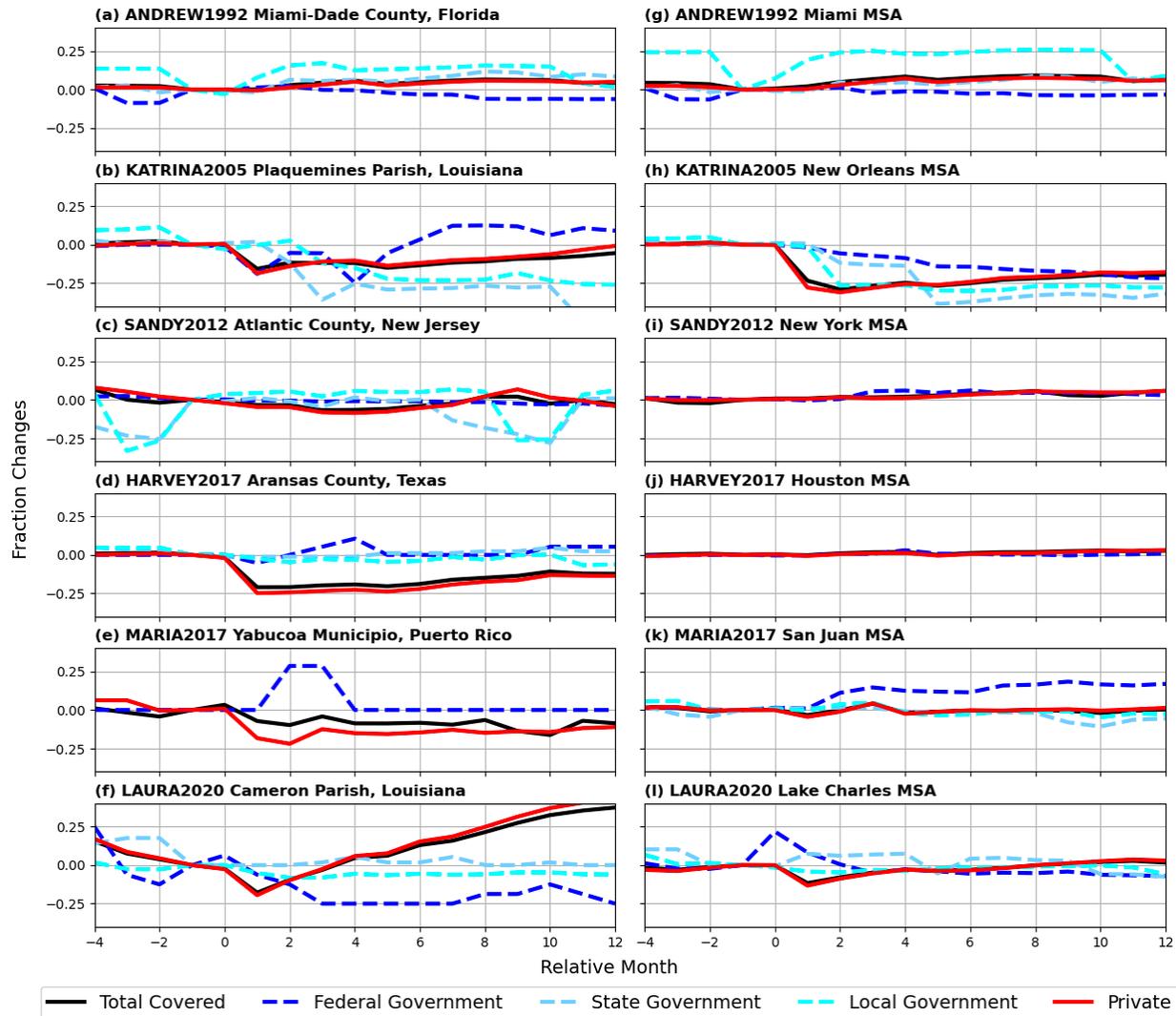

**Figure 2 Employment before and after hurricane impacts. To facilitate comparisons, the data are relative fractional changes with the Month -1 data as the reference. The nature of the analyzed employers follows the ownership code of the BLS data. As suggested by the figure legends, the black solid line indicates the total covered employment, and the red solid line shows the employment in the private sector. The dashed lines show the government**

**employment at Federal (deep blue), State (light blue), and Local (cyan) levels. The government sector information is missing in several panels due to incomplete data records. The horizontal axis indicates the month after hurricane passing (i.e., Month 0). The vertical axis indicates the relative, fractional change of employment after cyclone passing.**

Finally, the case analyses highlight that the monthly employment changes in storm-affected counties and MSAs are highly diverse. Although all the hurricanes caused heavy physical and economic damages, the employment changes after Hurricane Andrew (1992) (Figs. 2a and 2h) and in the other two MSA cases (Figs. 2i and 2j) appear minor or even slightly positive. The counterintuitive changes may be interpreted with different hypotheses. For example, the remarkable post-storm employment recovery in Cameron, Louisiana, which a main hub of liquefied natural gas (Fig. 2f), is likely attributable to the recent expansion of US natural gas export. On the other hand, other physical and socioeconomic factors may play some role in the diverse post-storm employment changes. We will explore some of these possibilities in the rest of Section 3 and ensuing sections.

## 3.2 Statistical Characteristics of More Diverse Samples

To explore to what extent the findings from the case studies may generalize, we examine all the US landfall storms during 1990–2020 (Fig. 3). Figure 3a shows that hurricanes and their remnants affect most US states east of the Rockies, with the highest frequency in the South and the Southeast. About a third of the examined counties experience hurricanes at least fifteen times over the three decades (Fig. 3b). The high-wind cases (>64 kt) mostly appear near the coast (Fig. 3c) and are rare in inland regions. This distribution is consistent with the fact that hurricanes tend to weaken after losing oceanic moisture support and experiencing land friction. In comparison, the storms with intense precipitation are less spatially concentrated (Fig. 3d), with many cases of heavy precipitation (>100 mm over 3 days) occurring inland. As suggested by Hurricane Ida (2021) and other storms (not shown), extreme precipitation associated with a storm can occur >1500 km away from the landfall point. The impacts may include widespread power outages and floods along the path of hurricane remnants. While these damages do not necessarily contribute to employment changes, they suggest employment changes associated with inland regions or extreme precipitation warrant further investigation.

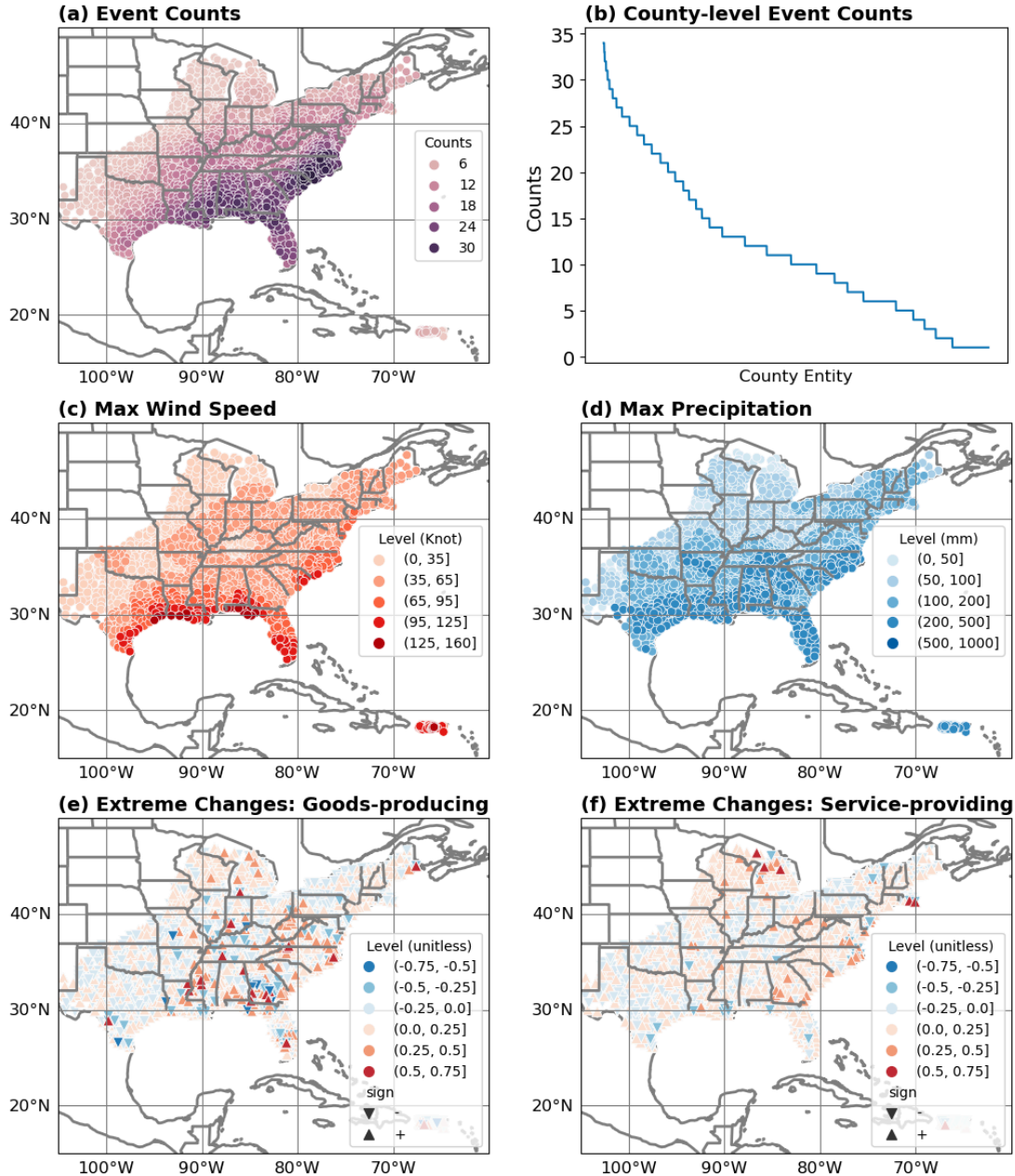

**Figure 3. Geographic distributions of storm hazards and related Month-1 employment changes. (a) The event counts at each county between 1990-2020, (b) county-level event counts sorted by values, (c) county-level wind maxima (kt), (d) county-level maximum 3-day precipitation (mm), (e) extreme fractional changes in goods-producing employment (unitless), (f) extreme fractional changes in service-proving employment (unitless). Each dot (a, c, d) or triangle (e, f) on the maps represents a county entry, with the largest values (c-d) or fractional changes (e-f) overlaid on the top of the plotting layer.**

We analyze employment changes in the goods-producing (Fig. 3e) and service-providing (Fig. 3f) industries separately. The primary motivation is the findings from previous US case studies (e.g., Groen *et al* 2020) and the study of non-US regions (e.g., Hsiang 2010) that suggest reconstruction activities can boost goods-producing industries (i.e., construction and materials demand). Figure 3e shows that some large changes (>25%) in goods-producing employment appear in the inland regions of Georgia and the Mississippi River basin, with many being employment gains. In comparison, large changes in the service-providing industries are generally negative on the coast of the GoM, most notably around New Orleans, LA. The negative GoM employment changes differ from the Atlantic coast, where more post-storm changes are positive. Overall, the descriptive analyses suggest post-storm employment changes are small for most inspected incidents. Large changes are uncommon but are not limited to the coastal region or wind hazards alone.

### 3.3 Conditioned Analyses of Hurricane Impacts

To explore what factors might affect post-storm employment changes, we first analyze employment changes conditioned on storm hazards and whether the affected region is in a coastal state (Fig. 4). While the post-storm employment changes are generally close to zero, a larger spread and a clearer negative skew emerge as storm hazards become more severe. The most distinct distribution is related to extreme wind ($\geq$96 kt), which corresponds to the threshold of Catgory-3 or major hurricanes. In those cases, a negative skewed distribution with a long tail appears in both the goods-producing and the service-providing industries. Assessments of the MSAs suggest similar employment sensitivity to major hurricanes (not shown). In coastal states, the employment changes associated with strong wind ($\geq$64 kt) and extreme precipitation ($\geq$150 mm over 3 days) are comparable. Compared to the unconditioned samples in the coastal states ("All"), the employment changes in the incidents with extreme precipitation show a weak negative skew (Fig. 4b). The negative skew becomes more apparent when extreme precipitation and strong wind happens together. The negative skew associated with extreme precipitation is also apparent for the incidents in the inland state (Fig. 4d), suggesting that hurricanes and their remnants may affect the inland employment.

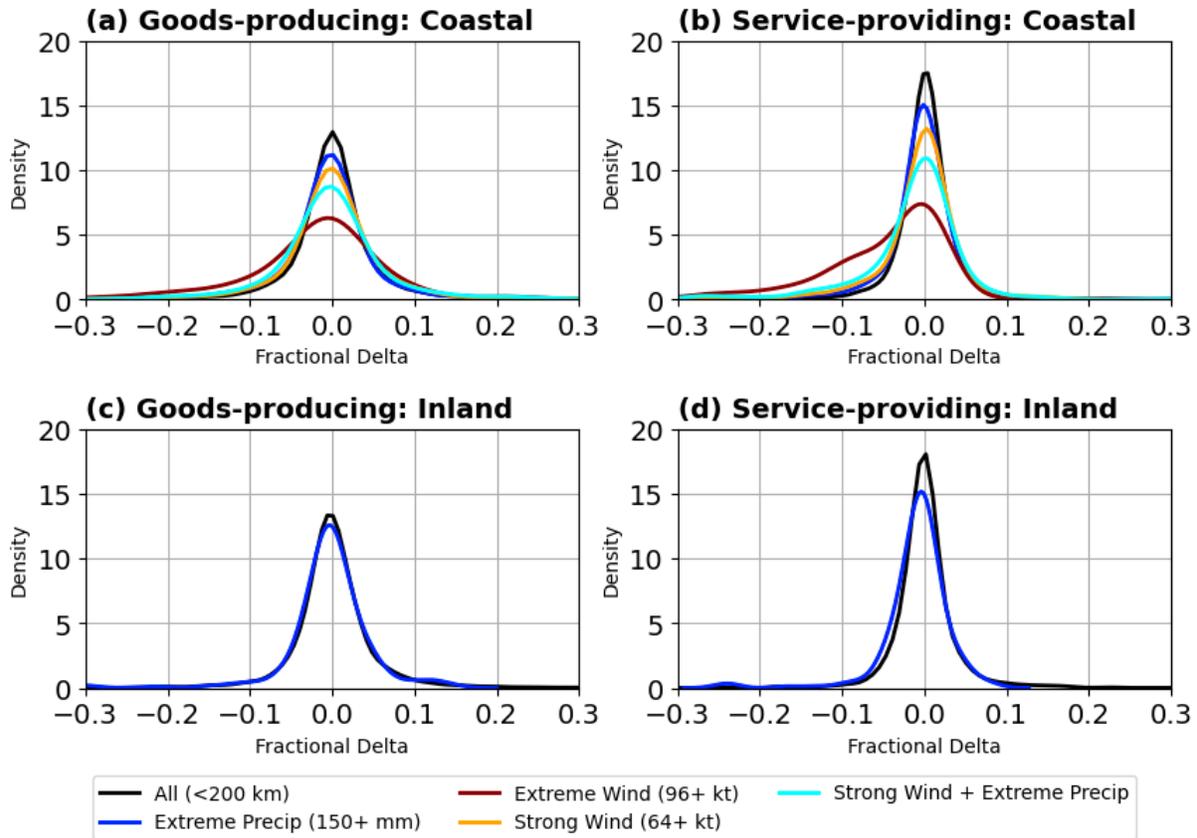

**Figure 4 Probability distribution functions of Month-1 employment changes. The changes in goods-producing and service-providing employment are analyzed separately for entities in (a, b) coastal states and (c, d) inland states. The color curves show the conditions of sample inputs, which include extreme wind (³96 kt), strong wind (³64 kt), and extreme precipitation (³150 mm over 3 days), and compound of strong wind and extreme precipitation. The thresholds of maximum wind speed correspond to Category-3 and Cateogry-1 hurricanes, respectively. Each of the analyzed groups have at least 200 samples. The samples of inland states do not include wind hazards as the sample size are small.**

Motivated by Figure 4 and previous studies (Hsiang 2010, Groen *et al* 2020), we extend the conditioned analysis to individual sectors and more time lags (Fig. 5). As a reference, the unconditioned samples suggest overall weak employment changes (Fig. 5a). In comparison, the same analyses revealed more pronounced changes after hurricanes with extreme wind (Fig. 5b). For goods-producing industries, the "construction" sector sees a 4% employment increase in Month 1 and an >20% increase by Month 6. This employment increase is accompanied by an 8% wage increase (not shown). This combination suggests the wage increase is driven by increasing labor demand rather than a sudden loss of labor forces associated with outmigration. This wage increase in the construction appears long-lasting and consistent with the econometric analyses by Tran and Wilson (2020). In comparison, the short-term effect of extreme wind on service-providing industries is overall negative, with leisure and hospitality suffering from the largest decrease (about -10%). This decrease might be related to business disruptions and temporary shifts in consumer spending priorities. Without seasonality adjustments, a full recovery to the pre-storm level happens by around Month 6 on average. Interestingly, the professional and business service sector starts showing a net increase after Month 6. The increase differs from the other service-providing sectors and is greater than the unconditioned reference samples (Fig. 5a), possibly indicating a delayed response to reconstruction activities. Lastly, we also examined the employment changes associated with strong wind ($\geq$64 kt) and extreme precipitation ($\geq$150 mm over 3 days) (not shown). These changes are comparable except that the growth of construction employment is more muted after the extreme precipitation incidents.

In summary, the descriptive analyses show that diverse employment changes occur after hurricane impacts. The changes appear to depend on storm hazards, industry sectors, and the affected entities. The findings indicate that findings on hurricane impacts on employment likely depend on the sample choice, which might help explain the spread in employment changes reported by previous studies. The analyses also provide useful hints and insights for further exploration (e.g., sensitivity in inland states). A caveat of the descriptive analyses is not accounting for the impacts of other factors that can affect the employment (e.g., economic cycles and socioeconomic covariates). While many post-storm changes are statistically different from zero (Fig. 5), the descriptive analyses alone cannot infer whether the changes are driven by storms or other factors. We will explore and validate some of the findings with more rigorous econometric analyses in Section 4.

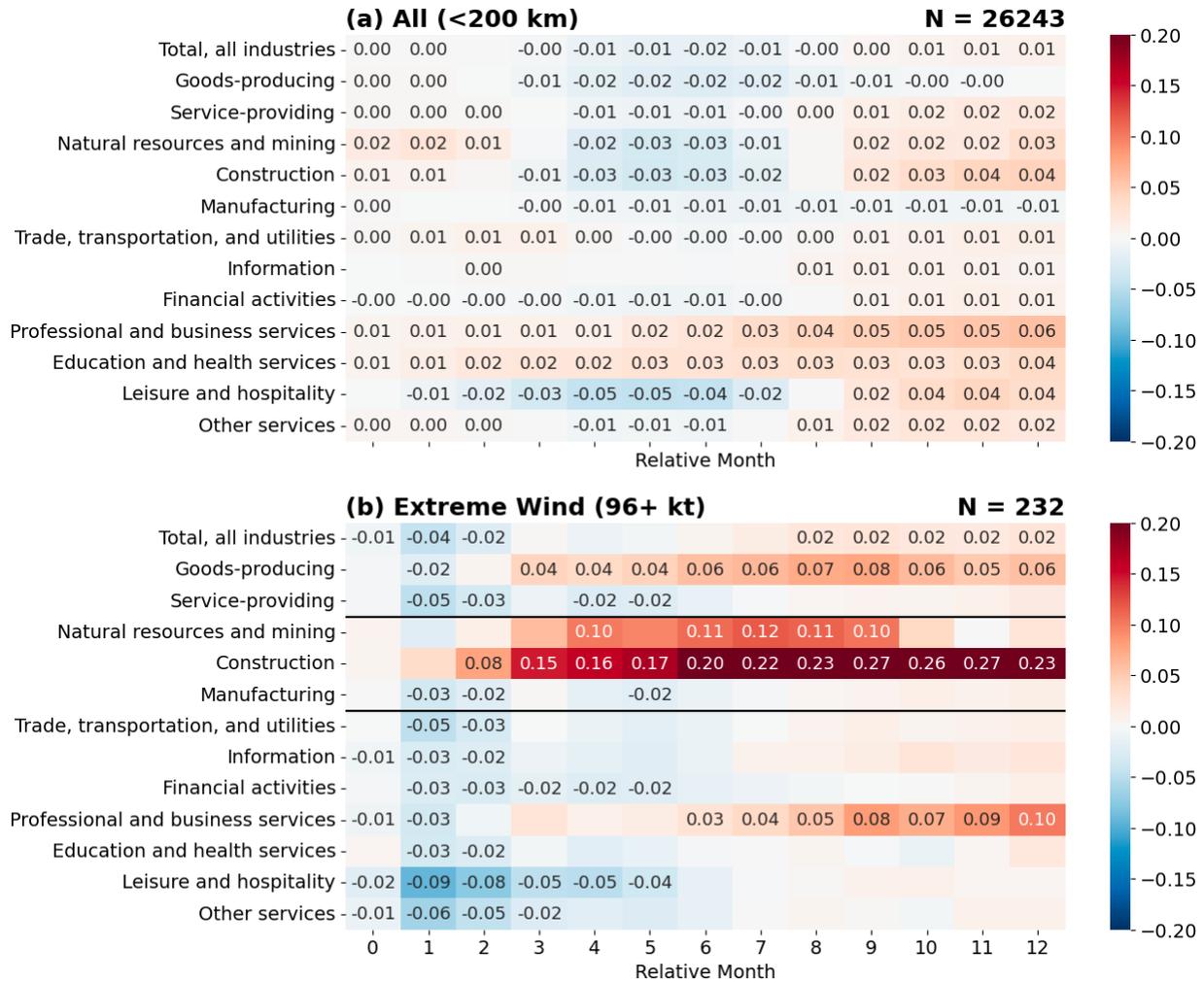

**Figure 5** Fractional employment changes in the North American Industry Classification (NAICS) subsectors after (a) any hurricanes and (b) hurricanes with extreme wind (96+ kt). The values in the cells denote mean values of employment changes, with the color shading showing the same values. Cells with numeric values displayed are significantly different from zero at the 99%-level. The significance test is a two-sided T-test with the null hypothesis that the employment changes are zero.

## 4. Econometric Analyses of Hurricane Impacts on Employment

The establish the causality between hurricanes and observed employment changes, this section will examine examples using econometric tools and more socioeconomic data. The additional data include the income per capita, the work-age population, and the education level (Section 2). We refrain from including additional physical variables, so the discussion can focus on the sensitivity of hurricane impact estimates to the choice of econometric tools. Before proceeding the discussion, we re-emphasize the samples analyzed in this section and the previous section differ moderately. This is because some socioeconomic data does not cover Puerto Rico, and the area coding of some county-level entities in Virginia is inconsistent among data sources. These issues, together with the inconsistent sampling frequency of socioeconomic data, may complicate the interpretation of the following econometric analyses. Nonetheless, we expect our findings to be at least qualitatively robust.

**4.1 Fixed Effects Model**

The fixed effects model attempt to control the effects of covariates and separate the hurricane impact on employment. We apply the model to private-sector employment in county-level entities and show the key model parameters in Table 2. The fixed effects model can fit the observed employment changes with statistically significant skills (p-value<0.05). The R2 overall suggests the fixed effects model explains only a moderate amount of variance, though the explaining power is higher when the industries are aggregated together (e.g., Total, all industries). Among the examined socioeconomic covariates, the income per capita consistently show a significant, positive relationship with employment. In comparison, the work-age population and education level show relationships with different significance and signs. The regression coefficient of the hurricane impact is consistently negative, suggesting negative, immediate impacts of hurricanes on employment. The statistical significance indicates the most robust impacts occur with the industry aggregations, as well as industries such as the construction and the leisure and hospitality.

**Table 2 Linear regression analysis of county-level employment ($log_{10}$). The analysis uses the fixed effects model of Eq. (1) with the demean pre-processing. The second column shows an indicator of the linear model's explaining power (R2 overall). The ensuing columns show the regression coefficients of income per capita (US Dollar), work-age population (person), education (percentage), and the hurricane impact dummy. The regression coefficients at the 0.05 significance level are in bold, while the insignificant values are in parentheses.**

| Employment (log10) | R2 overall | Income per capita | Work-age population | Education | Hurricane dummy |
|---|---|---|---|---|---|
| **Total, all industries** | 0.13 | **3.67E-06** | **4.97E-07** | **-3.83E-03** | **-4.90E-03** |
| **Goods-producing** | 0.12 | **6.42E-06** | (3.33E-08) | **-9.60E-03** | **-9.33E-03** |
| **Service-providing** | 0.07 | **1.92E-06** | **6.17E-07** | (6.33E-04) | **-3.46E-03** |
| **Natural resources and mining** | 0.03 | **5.20E-06** | **-4.68E-07** | **-8.19E-03** | **-1.09E-02** |
| **Construction** | 0.04 | **5.44E-06** | **2.99E-07** | **-1.80E-03** | **-1.70E-02** |
| **Manufacturing** | 0.07 | **4.39E-06** | (-6.90E-09) | **-1.20E-02** | (-3.08E-03) |
| **Trade, transportation, and utilities** | 0.05 | **2.07E-06** | **5.30E-07** | (-2.06E-04) | **-3.25E-03** |
| **Information** | 0.04 | **3.91E-06** | **6.62E-07** | (-1.59E-03) | (-2.52E-03) |
| **Financial activities** | 0.04 | **2.72E-06** | **5.71E-07** | (-5.05E-04) | (-1.64E-03) |
| **Professional and business services** | 0.01 | **3.10E-06** | **4.64E-07** | **3.84E-03** | **-2.88E-03** |
| **Education and health services** | 0.03 | **9.14E-07** | **6.97E-07** | (-9.43E-04) | (-7.05E-04) |
| **Leisure and hospitality** | 0.03 | **1.44E-06** | **7.24E-07** | **4.46E-03** | **-7.95E-03** |
| **Other Services** | 0.03 | **2.90E-06** | **4.89E-07** | (-1.38E-03) | (-2.73E-03) |

Compared to the descriptive analyses (Section 3), the negative impacts of hurricanes on employment appear more robust with the fixed effects model. Assuming impacts of the moderate sample differences are secondary, it is likely that controlling the covariates and considering the subject-specific effects help isolate hurricane impacts. While it is possible to modify or extend the fixed effects models (e.g., adding physical variables) to examine other interesting questions (Hsiang 2010, Strobl 2011), we leave this to future research and turn our attention to whether the estimated hurricane impacts are robust with other causality inference tools.

## 4.2 The Difference-in-Differences Analyses

Another common tool for the causality inference is the Difference-in-Differences (DiD) analyses, which can be considered as special case of fixed effects model. We follow the method outlined in Section 2.2.2 and examine employment changes before and after hurricanes. The pre-test of the parallel assumption suggests overall small differences between the hurricane-affected counties and the others. The significant differences in multi-industry aggregations (Figs. 6a–c) can be attributed to individual industries, such as natural resource and mining. While it is possible to reduce the small differences by refining the control groups, this effort will introduce additional assumptions and make the demonstrative analyses convoluted. For simplicity, we focus on comparing the Month-0 changes with the results from fixed effects model and briefly remark on the large, delayed changes.

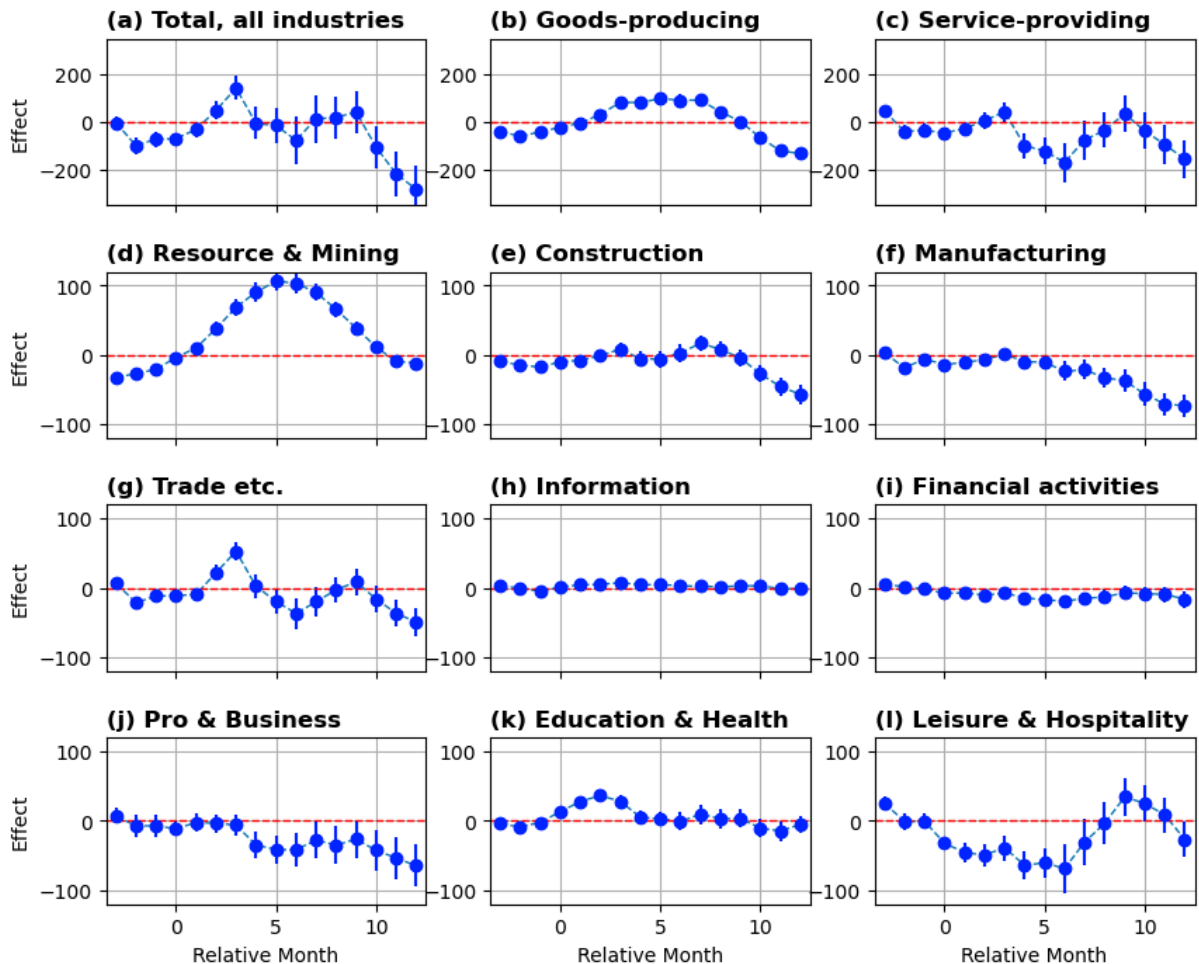

**Figure 6 The hurricane impacts on county-level employment estimated with the DiD analysis. The analyses examined the NAICS subsectors and used short titles for visual clarity. The subsector of other services shows small effects and is omitted for a tight figure layout. The blue dot indicates the average effects of treatment, and the error bar denotes the 95% confidence range.**

The DiD analysis suggests employment mostly decreases in the month of hurricane impacts. The exceptions like the education and health service generally correspond to the cases where the fixed effects model did not yield a significant relationship (Table 2). The DiD analysis also supports the significant, negative employment changes in leisure and hospitality after hurricane impacts. This relatively large employment loss is consistent the fixed effects model and the descriptive analyses (Section 3). Beyond Month 0, the large employment gains in natural resources and mining and the relatively muted changes in construction are surprising. The estimated effects are inconsistent with descriptive analyses or the expectation of more pronounced changes in construction employment. We speculate the employment decreases starting around Month 9 (Fig. 6) are related to the long-term, detrimental impacts of hurricane impacts reported by earlier studies (e.g., Strobl 2011, Hsiang and Jina 2014, Groen *et al* 2020). However, a satisfactory explanation about the timing of the negative signals and their later evolution remains elusive and warrant future research.

## 5. Data-Driven Analyses and Modeling of Employment Changes

Users of our new interdisciplinary dataset may also be interested in identifying potential contributors to post-storm employment changes and making skillful predictions about employment changes. For such tasks, the data pre-processing and ensuing modeling may involve subjective choices that depend on prior knowledge. For example, few economics studies explored how hurricane precipitation may affect the post-storm employment outcome, even though hurricane precipitation is significantly correlated with maximum wind speed and is projected to increase substantially in a warmer climate (e.g., Emanuel 2017, Knutson *et al* 2020). To serve future research and motivating interdisciplinary exchanges, we provide illustrative examples of analyzing and predicting short-term employment changes. Specifically, we apply the principal component analysis (PCA) to Month-1 employment changes to dissect scenarios of hurricane impacts. We also highlight the predictive power of nonlinear regression models and discuss potential applications involving climate change.

**5.1 Principal Component Analysis**

The abundance of samples affords an opportunity to conduct an agnostic, data-driven analysis of the storm-employment relationship. Focusing on the hurricanes with strong wind (>64 kt), we apply the PCA to the variables from the Best Track dataset and the BLS employment data. This choice helps us include a relatively large number of strong storms (N=1485) while including Puerto Rico and Virginia. For the sake of brevity, we refrain from discussing other subsets conditioned on regions or hazards. But using the open-source research code provided with this study, it is likely trivial for interested users to analyze a subset or incorporate other socioeconomic variables. Due to the spatial distribution of hurricane wind (Fig. 3), nearly all the analyzed samples are in coastal states. The analysis yields four leading modes that explain about 20.2%, 17.2%, 13.4%, and 10.7% of the variance of the standardized input data (Fig. 7). We will focus on three modes (PC1, PC2, PC4) that have clear loading in the employment changes. To help interpret each mode, we select observations with extreme weights and present the averages of two opposite groups (<10$^{th}$ percentile or >90$^{th}$ percentile) (Fig. 8).

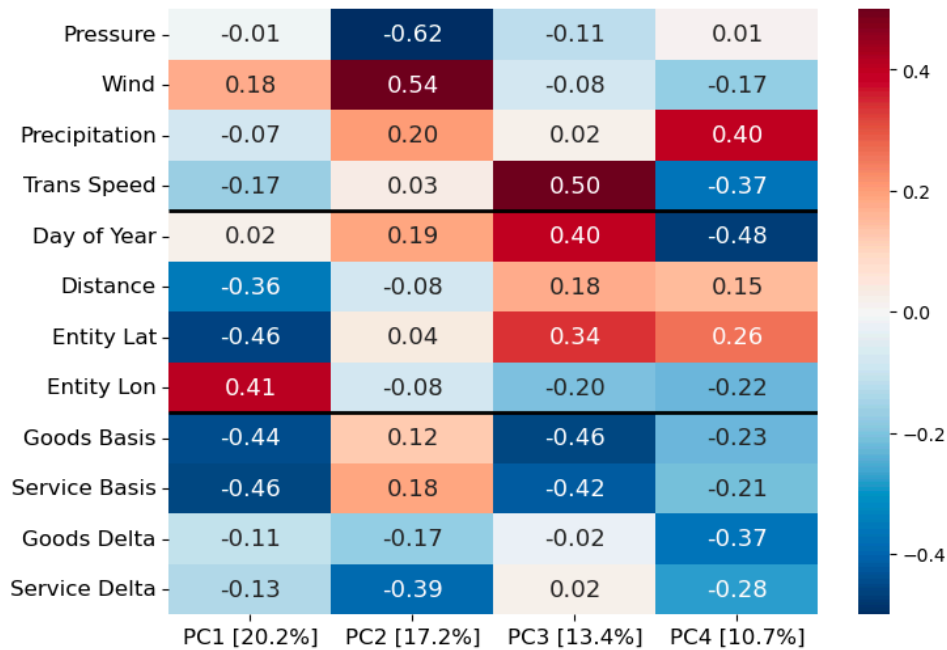

**Figure 7** The first four principal components (PCs) in the storm characteristics, spatial-temporal information, and employment basis and changes. The analyzed cases are hurricanes with strong winds (64+ kt). The numbers in the x-axis labels are the variance explained by the corresponding PCs. The employment basis data are pre-processed with log10, and each variable is standardized before PCA.

PC1 mainly characterizes a contrast between the US mainland and Puerto Rico (Figs. 8a–b vs Figs. 8c–d). The county-level entities in Puerto Rico are located at lower latitudes and employ fewer workers compared to their counterparts in the contiguous US. The hurricanes associated with PC1 tend to be stronger in Puerto Rico and move more slowly (Fig. 7). Perhaps unsurprisingly, the post-storm employment changes in Puerto Rico are more likely to skew negative. Interestingly, other Caribbean entities also show post-storm losses in employment (Mohan and Strobl 2021) and economic output (Hsiang 2010). As suggested by Hurricane Maria (2017), this sensitivity might be related to vulnerable infrastructure (e.g., power grid) and post-storm isolation due to shipping disruptions. Other contributors might include the small size of these labor markets and their dependence on service-providing industries (e.g., tourism). These possibilities may be key for understanding societal vulnerabilities and warrant future research.

PC2 represents the extensively-studied relationship between hurricane winds and post-storm employment changes (Figs. 7 and 8e–h). Consistent with past studies and the previous analyses (Sections 3), PC2 suggests hurricanes with more intense wind tend to be associated with greater employment losses, particularly in the service-providing sector. These intense storms also feature more extreme precipitation, which may execrate the impacts of damaging wind (Fig. 4). The more impactful hurricanes associated with PC2 also tend to occur later in the hurricane season and affect coastal entities with relatively large employment bases (Fig. 7). Examples of this group include Hurricane Katrina (2005) and several other high-impact cases that affected the GoM and the Atlantic coasts. While PC2 mostly reconfirm the existing knowledge, the discovery process does not rely on the knowledge from past literature. This independent reconfirmation lends credibility to the data-driven method and indicates it may provide new insights.

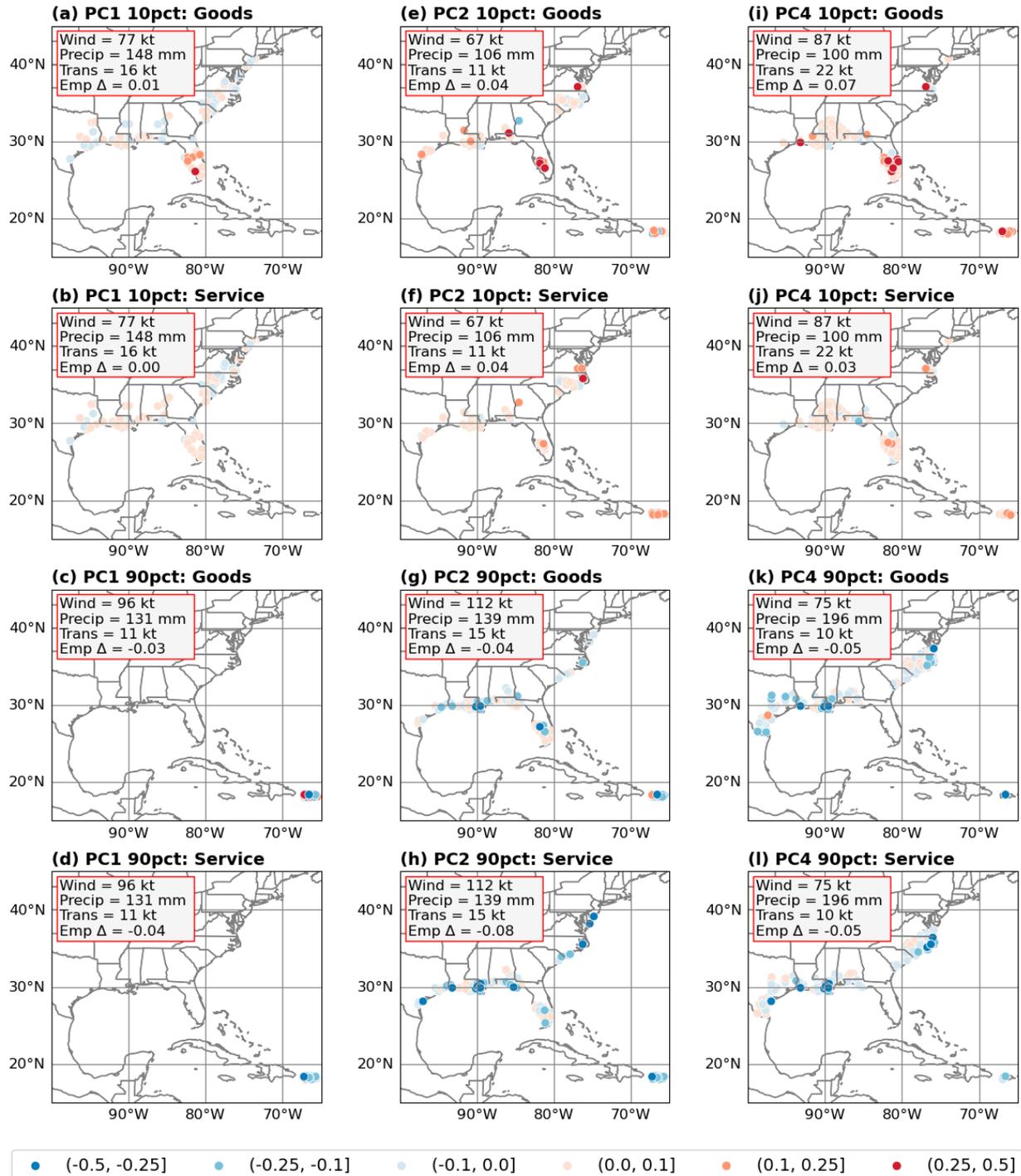

**Figure 8** Post-storm (Month-1) employment changes associated with PCs. (a) The goods-producing employment changes with the lowest values of PC1 (<10th percentile). (b) The service-providing employment changes with the lowest values of PC1 (<10th percentile). (c) The goods-producing employment changes with the highest values of PC1 (>90th percentile). (d) The service-providing employment changes with the highest values of PC1 (>90th

percentile). (e–h) Same as (a–d), but for PC2. (i–l) Same as (a–d), but for PC4. The employment changes are represented in unitless fractional values. The boxes with red borders show the average values of maximum wind speed, 3-day precipitation, translation speed of hurricanes, and employment changes in the corresponding subplots.**

PC4 suggests the potential of data-driven methods by characterizing a novel relationship between employment changes and hurricane precipitation (Figs. 7 and 8i–l). The mode is consistent with the physical relationship between hurricane precipitation and translation speed. The relationship is a key contributor to the record-breaking rainfall of slow-moving Hurricane Harvey (2017), which is followed by employment losses (Figure 2d). A composite analysis suggests that a group of wet, weak, and slow-moving storms in the early hurricane season are associated employment losses in the GoM and the mid-Atlantic states. On the other hand, the South Florida tend to experience an employment increase after late-season, fast-moving, and relatively dry storms. A closer inspection of industries suggests that the leading contributor to this employment increase is the natural resources and mining sector. Interestingly, Figures 8a, e, and i suggest the goods-producing employment in South Florida tends to increase after weak hurricanes. Whether this is a robust feature with unique socioeconomic drivers warrants future research.

Overall, the PCA analyses demonstrate the potential of combining more data and existing analytical tools in studying associations between employment changes and diverse factors. Although the PCA does not directly serve the causal inference or predictive modeling, it can help highlight that the covarying socioeconomic and physical factors in various scenarios. The findings may motivate more rigorous studies or real-world applications.

### 5.2 Predictive Modeling of Post-storm Employment Change

For applications such as expediting aids and services, accurate predictive modeling of post-storm employment shocks is valuable. While past studies often rely on linear models, one may consider nonlinear models to safeguard against questionable model assumptions. We construct a simple case to demonstrate a possible workflow of predictive modeling and the potential benefits of using a nonlinear model. In this simple case, the post-storm (Month-1) employment changes are modeled with the multiple linear regression (MLR) model and the Random Forest (RF) model. The modeling task is to use hurricane information and the basic information of affected counties to predict employment changes. Details about the training (e.g., dropping colinear variables) and cross-validation are available in Supplementary Information. We emphasize that the analyses here are not a causal inference practice. The demonstration does not seek to fully incorporate the latest development of machine learning practice in social sciences (e.g., Chernozhukov *et al* 2018, Athey and Wager 2019), either. The rest of this section will focus on the model performance and result interpretation of demonstrative cases.

**Table 3 Skills of the linear regression (LR) and the Random Forests (RF) in predicting the service-providing employment changes one month after storms. R2 stands for coefficient of determination, and MAE represents mean absolute error. The standard deviation of skill metrics is estimated with the five-fold cross-validation with the K-fold strategy. The employment basis data are pre-processed with log10, and each variable is standardized before regressions.**

|  | All (<200 km) | Extreme Precipitation (>150 mm) | Strong Wind (>64 kt) | Extreme Wind (>96 kt) |
|---|---|---|---|---|
| **Sample Num** | N=26243 | N=2060 | N=1485 | N=232 |
| **LR R2** | 0.03±0.01 | 0.04±0.02 | 0.12±0.07 | 0.15±0.07 |
| **LR MAE** | 0.03±0.00 | 0.03±0.00 | 0.03±0.00 | 0.05±0.00 |
| **RF R2** | 0.19±0.02 | 0.21±0.08 | 0.27±0.09 | 0.35±0.07 |
| **RF MAE** | 0.02±0.00 | 0.03±0.00 | 0.03±0.00 | 0.04±0.00 |

The MLR and the RF models perform better in predicting employment changes in the service-providing sector than in the good-producing sector (Supplementary Fig. 3). This observation holds in all the experiments with or without conditioning on storm hazards (not shown). The result is perhaps unsurprising as Month-1 employment changes in the service-providing sector show strong correlations with storm metrics (Supplementary Fig. 2). Given the relatively poor predictive skills with the goods-producing employment, the ensuing discussion will focus on the service-provide employment. Table 3 shows that the MLR and the RF models can predict post-storm employment changes with skills, especially for storms associated with more severe hazards. The prediction skill depends on the model choice, with the RF model showing superior skills for every sample group. The higher skills are attained with test samples in a five-fold cross-validation and unlikely a result of overfitting training data (Chernozhukov *et al* 2018). The relatively good performance of the RF model is perhaps unsurprising since the linear model assumption would be questionable if the unknown functional form of employment changes is nonlinear.

Although an RF model does not present an analytical formula, it can provide useful estimates of feature importance in the regression (Fig. 10). When considering all the storm cases, the most important feature is the time of hurricane landfall. The sensitivity was detected by PCA, with PC2 and PC4 showing region-dependent associations between landfall time and employment changes (Section 5.1). The next three features that explain the model skill are the day of entity latitudes, longitudes, and employment basis (Fig. 10a). These features explain about 46% of the RF models' predictive power, suggesting the RF may have learned subject-specific sensitivities. For example, the geographic location of a county-level entity may implicitly contain information (e.g., inland versus near-coast exposure) that can be leveraged by decision trees. Lastly, the physical features of storms (i.e., wind, precipitation, and translation speed) are secondary contributors and collectively explain 24% predictive power of the RF model. When applied to intense storms (>64 kt; Fig. 10b), the feature importance analysis suggests the feature weights of the RF regression model are sensitive to the input sample. Compared to the unconditioned samples (Fig. 10a), the predictive power associated with the landfall time dropped to near the bottom, while the importance of wind hazard and the storm-entity distance increases relative to other variables. The higher sensitivity is consistent with the destructive wind near the center of strong hurricanes.

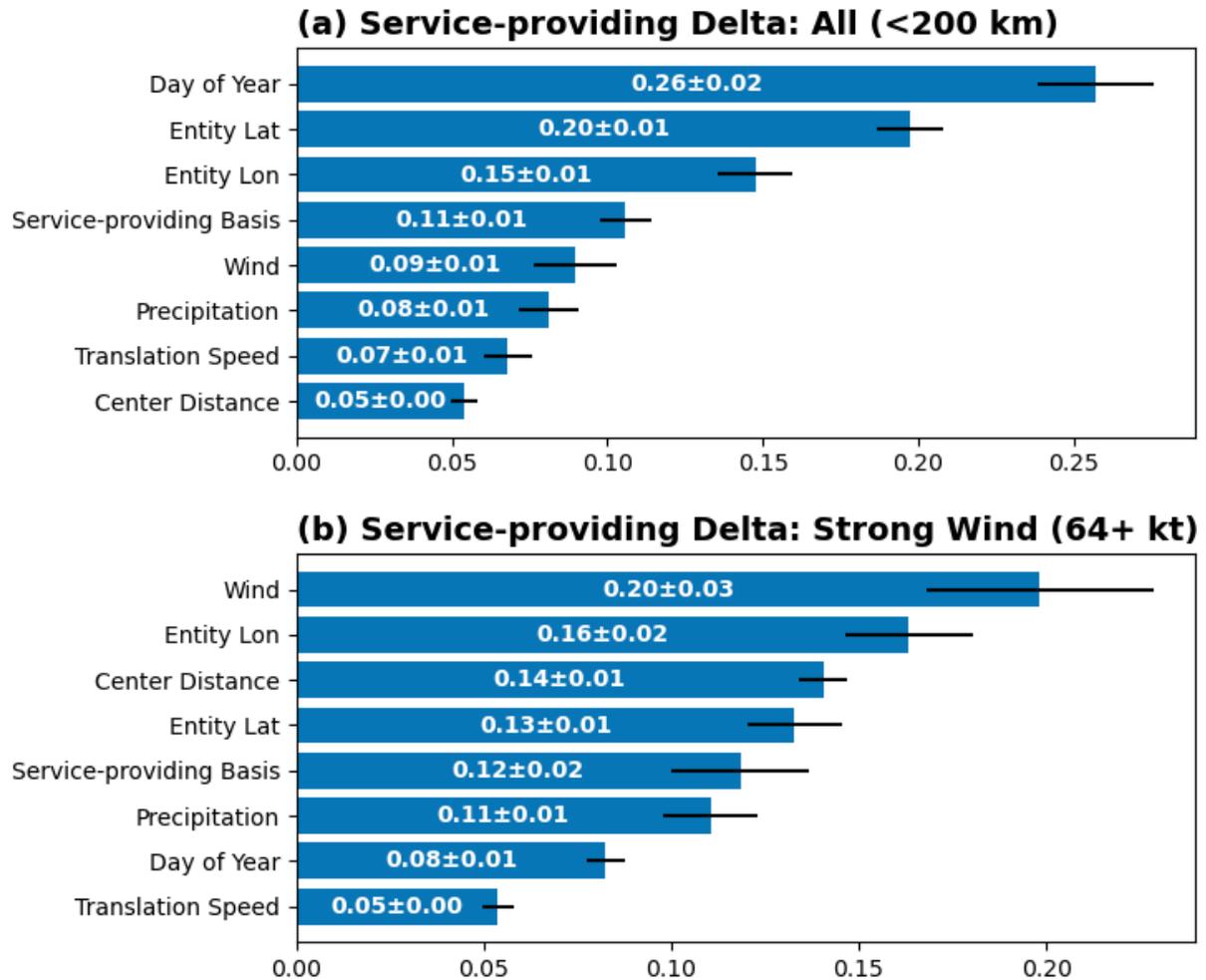

**Figure 9 Feature importance estimated from the RF models for the Month-1 servicing-providing employment. (a) All the overland storms; (b) the overland storms with >64 kt wind. The horizontal black lines indicate the range of ±1 standard deviation. The bar chart values are shown in the format of mean ± standard deviation based on the cross-validation results. The feature importance is estimated based on the statistics of decision trees.**

The predictive skill of the RF model suggests it might support future explorations such as testing the sensitivity of post-storm employment changes to different hazard scenarios. Assuming most hazards under consideration do not exceed the range of historical observations, one may adopt the RF model trained with historical cases (e.g., strong wind 64+ kt), apply hypothetical fractional changes to the hurricane hazards, and explore how these changes may affect the predicted employment changes. As an illustrative example, we examine how increases in wind speed and precipitation may affect the RF model predictions. The wind and precipitation hazards of hurricanes may intensify due to anthropogenic warming (Knutson *et al* 2020). Figure 10a suggest the RF prediction with the historical hurricanes is consistent the observed post-storm employment changes, though the predictions has a narrower distribution. In the scenarios with stronger wind and precipitation, the distributions shift farther towards negative values. While the mean changes are relatively small across the tested parameter space (Fig. 11b), the tail of negative post-storm employment changes (-0.10 to -0.05) becomes up to about three times (approximately 4K warming, red) fatter than the reference prediction (historical, blue). The lack of distribution changes beyond -0.10 may be related to fact that the RF model underestimates the most extreme employment losses. For example, an experiment with the six cases in Fig. 2 suggests the RF model underestimate the group-mean losses in the service-providing employment substantially (-8.5% versus -15.5%). Given this weakness, we present this analysis as an estimate of lower bounds and leave more in-depth research to the future.

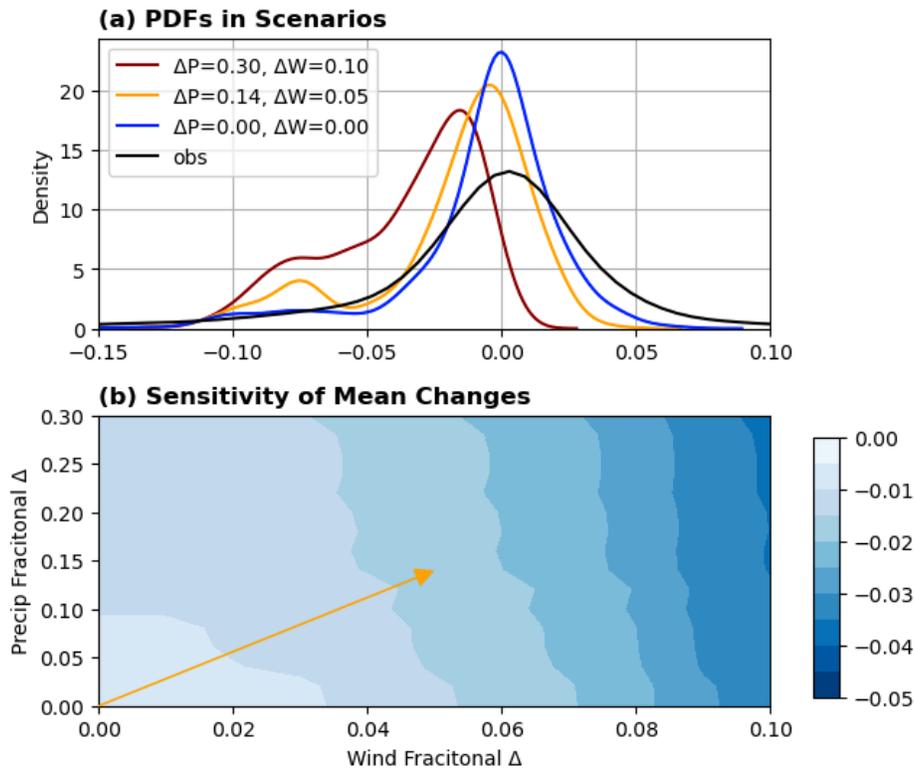

**Figure 10. Sensitivity of the RF model prediction of post-storm employment changes to increases in wind and precipitation. (a) Probability distribution functions of the observation**

and model predictions. The blue scenario is generated using the predictions with the information of observed strong wind (>64 kt) cases. The orange scenario is generated with scaled hurricane hazards, where precipitation increases by 14% and wind increases by 5%. The red scenario is the same except that the precipitation increases by 28% and wind increases by 10%. (b) The sensitivity of the mean post-storm employment changes predicted by the RF model to different combinations of hazard changes. The arrow highlights the orange scenario in (a), which is motivated by physical studies of global TC activity. Specifically, the consensus about 2-K global warming suggested 1) the maximum wind of global TCs would increase by about 5%; and 2) the storm precipitation would increase by about 14% (Knutson et al. 2020). The most extreme warming in the twenty-first century may reach 4 K (SSP5-8.5), and the projected hazard changes roughly correspond to the red scenario.

## 6. Summary and Discussion

This study presents a new open-access dataset that covers the US employment, selected socioeconomic data, and hurricane hazards during 1990–2020. Compared to previous studies with limited cases or regions, the new dataset may contribute to a more comprehensive understanding of the post-storm employment changes in the US. Our demonstrative analyses include descriptive analyses, econometric analyses, and data-driven analyses. The analyses mainly cover the county-level entities and include granular information about employment sectors. The findings include:

- Large employment changes can occur after storm impacts, especially in the private sector. Regarding the initial employment shock, some storms with extreme wind (>96 kt) can reduce local employment by >30%. Nonetheless, the observed employment changes after storms are close to zero on average.
- Post-storm employment changes depend on individual sectors and hazard severity. After major hurricanes (>96 kt), leisure and hospitality employment on average decreases by about 10% in Month 1. The construction employment increases for an extended period (>25% in Month 9). Several other sectors show more moderate but statistically significant changes.

- A simple fixed effects model and a DiD analysis suggest the immediate impact of hurricanes on the US county-level employment are overall negative. While the DiD analysis suggests a multi-month decrease in the employment of leisure and hospitality, the analysis does not show a pronounced increase in construction employment that are observed with strong hurricanes.
- Descriptive analyses and the PCA suggest hurricane-related extreme precipitation precedes decreases in service-providing employment. The extreme precipitation received limited attention by past hurricane-employment studies and is likely important for potential employment responses in inland regions. The PCA analysis also highlight the vulnerability of Puerto Rico.
- The Month-1 employment changes can be skillfully predicted based on storm features and county-level information. The prediction is more skillful for the servicing-providing industries and more severe storms. The RF model outperforms a baseline MLR model and may be useful for exploring sensitivities of employment to the intensifying hurricane hazards in a changing climate.

The findings complement existing studies and have implications for future studies of how hurricanes affect employment and economies. For example, our analysis highlights that service-providing employment tends to respond negatively to hurricanes. A plausible hypothesis is that regional economies with a greater dependence on service-providing industries (e.g., tourist destinations) might be more vulnerable to adverse storm impacts. Meanwhile, the PCA suggests that employment in Puerto Rico is particularly vulnerable to adverse storm impacts, which might be representative of other isolated tropical islands. With our new dataset, it is likely more tangible for future research to consolidate the modeling of short-term employment changes and long-term economic impacts (e.g., Hsiang 2010, Strobl 2011, Hsiang and Jina 2014) to create a seamless understanding of how hurricanes may affect the well-being of communities and individuals. Finally, the severity of hurricane hazards (e.g., maximum wind and precipitation) is projected to increase (Knutson *et al* 2020, Seneviratne *et al* 2021). An exploratory sensitivity test with the RF model suggests the projected hurricane changes may contribute to more extreme short-term employment losses. To build resilient economies and protect vulnerable populations, the research of physical changes and socioeconomic changes (e.g., coastal growth) need to converge to delineate emerging risks.

The study has several limitations that arise from practical considerations. First, the extraction of storm hazard information involves substantial simplifications, such as ignoring the geospatial heterogeneity of hazards within individual storms. Future studies could consider using climate reanalysis datasets to derive a refined view of the county-level exposure to storm hazards. Second, the predictive modeling here serves as a minimal example but makes little effort to achieve the best skill. Future skill improvements may be possible by incorporating more predictor features (e.g., storm surge) and tuning model algorithms. Third, the study focuses on the US data, its conclusions should not be extended without scrutiny. For example, infrastructure standards are generally weaker in developing countries, so weaker storms might incur more substantial impacts. Lastly, while our simple benchmark provides encouraging evidence for implementing nonlinear machine learning models for predictive modeling, a well-designed MLR model likely can outperform our baseline MLR model and deliver respectable performance.

Despite these limitations, this study is likely a meaningful step toward better understanding the socioeconomic impacts of landfall hurricanes. A natural way to extend this study is to conduct field investigations to verify the analytical findings and pursue a more granular understanding of such populations. Given that storm characteristics can be skillfully predicted with modern weather models (Alaka *et al* 2024), it is plausible to predict hurricane impacts on employment *before* landfall. Such applications can be consolidated with other new tools that forecast impacts of other weather-climate hazards (e.g., Alipour *et al* 2020). The consolidated system may help target the most vulnerable by enabling more timely, effective aid programs. We invite researchers and other users to consider leveraging our work (e.g., dataset and code) to explore hurricane impacts and help communities prepare for challenges in a changing climate.

**Acknowledgment**

The authors thank the Bureau of Labor Statistics (BLS) and the National Centers for Environmental Information (NCEI) of the US for making the research data publicly available. Open-source Python software including Numpy (Harris *et al* 2020), Pandas (McKinney 2010), Xarray (Hoyer and Hamman 2017), Scikit-learn (Pedregosa *et al* 2012), linearmodels (https://bashtage.github.io/linearmodels/), statsmodel (Seabold and Perktold 2010), seaborn (Waskom 2021), matplotlib (Hunter 2007), and cartopy (Met Office 2010) greatly facilitated the analysis. We gratefully acknowledge Dr. Tatyana Deryugina and multiple anonymous reviewers for insightful input that helps improve this manuscript substantially. GZ is supported by the faculty development fund of the University of Illinois at Urbana-Champaign and the US National Science Foundation (Award 2327959).

**Data Availability Statement**

The input data are available online and publicly accessible. We list the dataset and links as follows:

- The employment data (https://www.bls.gov/cew/downloadable-data-files.htm)
- The personal income data (https://apps.bea.gov/regional/downloadzip.cfm)
- The county-level education data (https://www.ers.usda.gov/data-products/county-level-data-sets/county-level-data-sets-download-data/)
- The county-level population data (https://seer.cancer.gov/popdata/download.html)
- The hurricane data (https://www.ncei.noaa.gov/products/international-best-track-archive)
- The precipitation data (https://downloads.psl.noaa.gov/Datasets/cpc_global_precip/).

The software to remove the seasonal cycle from employment data is available on the US Census website (https://www.census.gov/data/software/x13as.html). The research code was developed with open-source Python libraries. The consolidated dataset and analysis scripts will be available via Zenodo upon the acceptance decision. We publish intermediate data but not the raw data from the aforementioned parties. Users shall comply with the original licenses and terms of the relevant parties that provided the data and software.

# References

Alaka G J, Sippel J A, Zhang Z, Kim H-S, Marks F D, Tallapragada V, Mehra A, Zhang X, Poyer A and Gopalakrishnan S G 2024 Lifetime Performance of the Operational Hurricane Weather Research and Forecasting (HWRF) Model for North Atlantic Tropical Cyclones *Bulletin of the American Meteorological Society* Online: https://journals.ametsoc.org/view/journals/bams/aop/BAMS-D-23-0139.1/BAMS-D-23-0139.1.xml

Alipour A, Ahmadalipour A, Abbaszadeh P and Moradkhani H 2020 Leveraging machine learning for predicting flash flood damage in the Southeast US *Environ. Res. Lett.* **15** 024011

Athey S and Wager S 2019 Estimating Treatment Effects with Causal Forests: An Application *Observational Studies* **5** 37–51

Bach P, Chernozhukov V, Kurz M S and Spindler M 2022 DoubleML - An Object-Oriented Implementation of Double Machine Learning in Python *Journal of Machine Learning Research* **23** 1–6

Bartik A W, Bertrand M, Cullen Z, Glaeser E L, Luca M and Stanton C 2020 The impact of COVID-19 on small business outcomes and expectations *Proc. Natl. Acad. Sci. U.S.A.* **117** 17656–66

Belasen A R and Polachek S W 2008 How Hurricanes Affect Wages and Employment in Local Labor Markets *American Economic Review* **98** 49–53

Breiman L 2001 Random Forests *Machine Learning* **45** 5–32

Brown S P, Mason S L and Tille R B 2006 The effect of Hurricane Katrina on employment and unemployment *Monthly Labor Review* 52

Chen M, Shi W, Xie P, Silva V B S, Kousky V E, Wayne Higgins R and Janowiak J E 2008 Assessing objective techniques for gauge-based analyses of global daily precipitation *J. Geophys. Res.* **113** D04110

Chernozhukov V, Chetverikov D, Demirer M, Duflo E, Hansen C, Newey W and Robins J 2018 Double/debiased machine learning for treatment and structural parameters *The Econometrics Journal* **21** C1–68

Deryugina T 2017 The Fiscal Cost of Hurricanes: Disaster Aid versus Social Insurance *American Economic Journal: Economic Policy* **9** 168–98

Deryugina T, Kawano L and Levitt S 2018 The Economic Impact of Hurricane Katrina on Its Victims: Evidence from Individual Tax Returns *American Economic Journal: Applied Economics* **10** 202–33

Emanuel K 2017 Assessing the present and future probability of Hurricane Harvey's rainfall *Proc. Natl. Acad. Sci. U.S.A.* **114** 12681–4

Evans C, Wood K M, Aberson S D, Archambault H M, Milrad S M, Bosart L F, Corbosiero K L, Davis C A, Dias Pinto J R, Doyle J, Fogarty C, Galarneau T J, Grams C M, Griffin K S, Gyakum J, Hart R E, Kitabatake N, Lentink H S, McTaggart-Cowan R, Perrie W, Quinting J F D, Reynolds C A, Riemer M, Ritchie E A, Sun Y and Zhang F 2017 The Extratropical Transition of Tropical Cyclones. Part I: Cyclone Evolution and Direct Impacts *Mon. Wea. Rev.* **145** 4317–44

FEMA Effects of Disasters on Small Businesses Online: https://emilms.fema.gov/is_0111a/groups/23.html

Galea S, Brewin C R, Gruber M, Jones R T, King D W, King L A, McNally R J, Ursano R J, Petukhova M and Kessler R C 2007 Exposure to Hurricane-Related Stressors and Mental Illness After Hurricane Katrina *Arch Gen Psychiatry* **64** 1427

Garber M, Unger L, White J and Wohlford L 2006 Hurricane Katrina's effects on industry employment and wages *Monthly Labor Review* **129** 22–39

Goulbourne R D 2021 *An Analysis of the Effect of Hurricanes on Economic Growth and Labor Market Outcomes* thesis (University of Alabama Libraries) Online: https://ir.ua.edu/handle/123456789/8155

Groen J A, Kutzbach M J and Polivka A E 2020 Storms and Jobs: The Effect of Hurricanes on Individuals' Employment and Earnings over the Long Term *Journal of Labor Economics* **38** 653–85

Groen J A and Polivka A E 2008 The Effect of Hurricane Katrina on the Labor Market Outcomes of Evacuees *American Economic Review* **98** 43–8

Harris C R, Millman K J, Van Der Walt S J, Gommers R, Virtanen P, Cournapeau D, Wieser E, Taylor J, Berg S, Smith N J, Kern R, Picus M, Hoyer S, Van Kerkwijk M H, Brett M, Haldane A, Del Río J F, Wiebe M, Peterson P, Gérard-Marchant P, Sheppard K, Reddy T, Weckesser W, Abbasi H, Gohlke C and Oliphant T E 2020 Array programming with NumPy *Nature* **585** 357–62

Hemmati M, Camargo S J and Sobel A H 2022 How are Atlantic basin-wide hurricane activity and economic losses related? *Environ. Res.: Climate* **1** 021002

Hereid K A 2022 Hurricane Risk Management Strategies for Insurers in a Changing Climate *Hurricane Risk in a Changing Climate* Hurricane Risk vol 2, ed J M Collins and J M Done (Cham: Springer International Publishing) pp 1–23 Online: https://link.springer.com/10.1007/978-3-031-08568-0_1

Hiti M, Mills, C K and Sarkar A 2022 Small Business Recovery after Natural Disasters *Liberty Street Economics* Online: https://libertystreeteconomics.newyorkfed.org/2022/09/small-business-recovery-after-natural-disasters/
37

Hoyer S and Hamman J 2017 xarray: N-D labeled Arrays and Datasets in Python *JORS* **5** 10

Hsiang S and Jina A 2014 *The Causal Effect of Environmental Catastrophe on Long-Run Economic Growth: Evidence From 6,700 Cyclones* (Cambridge, MA: National Bureau of Economic Research) Online: http://www.nber.org/papers/w20352.pdf

Hsiang S M 2010 Temperatures and cyclones strongly associated with economic production in the Caribbean and Central America *Proc. Natl. Acad. Sci. U.S.A.* **107** 15367–72

Hunter J D 2007 Matplotlib: A 2D Graphics Environment *Comput. Sci. Eng.* **9** 90–5

Jerch R, Kahn M E and Lin G C 2023 Local public finance dynamics and hurricane shocks *Journal of Urban Economics* **134** 103516

Jolliffe I T and Cadima J 2016 Principal component analysis: a review and recent developments *Phil. Trans. R. Soc. A.* **374** 20150202

Knapp K R, Kruk M C, Levinson D H, Diamond H J and Neumann C J 2010 The International Best Track Archive for Climate Stewardship (IBTrACS): Unifying Tropical Cyclone Data *Bull. Amer. Meteor. Soc.* **91** 363–76

Knutson T, Camargo S J, Chan J C L, Emanuel K, Ho C-H, Kossin J, Mohapatra M, Satoh M, Sugi M, Walsh K and Wu L 2020 Tropical Cyclones and Climate Change Assessment: Part II: Projected Response to Anthropogenic Warming *Bulletin of the American Meteorological Society* **101** E303–22

Kossin J P 2018 A global slowdown of tropical-cyclone translation speed *Nature* **558** 104–7

Kossin J P, Emanuel K A and Vecchi G A 2014 The poleward migration of the location of tropical cyclone maximum intensity *Nature* **509** 349–52

McIntosh M F 2008 Measuring the Labor Market Impacts of Hurricane Katrina Migration: Evidence from Houston, Texas *American Economic Review* **98** 54–7

McKinney W 2010 Data structures for statistical computing in Python. *SciPy* vol 445 pp 51–6

Met Office 2010 *Cartopy: a cartographic python library with a Matplotlib interface* (Exeter, Devon) Online: https://scitools.org.uk/cartopy

Mohan P and Strobl E 2021 Hurricanes and their implications for unemployment: Evidence from the Caribbean *ILO Working Paper* **26** Online: https://www.ilo.org/legacy/english/intserv/working-papers/wp026/index.html

Nordhaus W D 2010 The Economics of Hurricanes and Implications of Global Warming *Clim. Change Econ.* **01** 1–20

Parks R M, Anderson G B, Nethery R C, Navas-Acien A, Dominici F and Kioumourtzoglou M-A 2021 Tropical cyclone exposure is associated with increased hospitalization rates in older adults *Nat Commun* **12** 1545

Pedregosa F, Varoquaux G, Gramfort A, Michel V, Thirion B, Grisel O, Blondel M, Müller A, Nothman J, Louppe G, Prettenhofer P, Weiss R, Dubourg V, Vanderplas J, Passos A, Cournapeau D, Brucher M, Perrot M and Duchesnay É 2012 Scikit-learn: Machine Learning in Python Online: https://arxiv.org/abs/1201.0490

Rhodes J, Chan C, Paxson C, Rouse C E, Waters M and Fussell E 2010 The impact of Hurricane Katrina on the mental and physical health of low-income parents in New Orleans. *American Journal of Orthopsychiatry* **80** 237–47

Seabold S and Perktold J 2010 statsmodels: Econometric and statistical modeling with python *9th Python in Science Conference*

Seneviratne S I, Zhang X, Adnan M, Badi W, Dereczynski C, Di Luca A, Ghosh S, Iskandar I, Kossin J, Lewis S, Otto F, Pinto I, Satoh M, Vicente-Serrano S M, Wehner M and Zhou B 2021 Weather and Climate Extreme Events in a Changing Climate *Climate Change 2021: The Physical Science Basis. Contribution of Working Group I to the Sixth Assessment Report of the Intergovernmental Panel on Climate Change* (Cambridge University Presss, Cambridge, United Kingdom and New York, NY, USA) pp 1513–766 Online: 10.1017/9781009157896.013

Smith A B 2020 U.S. Billion-dollar Weather and Climate Disasters, 1980 - present (NCEI Accession 0209268) Online: https://www.ncei.noaa.gov/archive/accession/0209268

Smith A B and Katz R W 2013 US billion-dollar weather and climate disasters: data sources, trends, accuracy and biases *Nat Hazards* **67** 387–410

Strobl E 2011 The Economic Growth Impact of Hurricanes: Evidence from U.S. Coastal Counties *Review of Economics and Statistics* **93** 575–89

Strobl E 2012 The economic growth impact of natural disasters in developing countries: Evidence from hurricane strikes in the Central American and Caribbean regions *Journal of Development Economics* **97** 130–41

Studholme J, Fedorov A V, Gulev S K, Emanuel K and Hodges K 2022 Poleward expansion of tropical cyclone latitudes in warming climates *Nat. Geosci.* **15** 14–28

Tran B R and Wilson D J 2020 The Local Economic Impact of Natural Disasters *ERWP* 1.000-61

Waskom M 2021 seaborn: statistical data visualization *JOSS* **6** 3021

Wilson S G and Fischetti T R 2010 *Coastline population trends in the United States 1960 to 2008* (US Department of Commerce, Economics and Statistics Administration, US …)

World Meteorological Organization 2021 *WMO Atlas of Mortality and Economic Losses from Weather, Climate and Water Extremes (1970–2019) (WMO-No. 1267)* (Geneva: WMO)

Wu X, Zhou L, Guo J and Liu H 2017 Impacts of Typhoons on Local Labor Markets based on GMM: An Empirical Study of Guangdong Province, China *Weather, Climate, and Society* **9** 255–66

Zhang G, Murakami H, Knutson T R, Mizuta R and Yoshida K 2020 Tropical cyclone motion in a changing climate *Sci. Adv.* **6** eaaz7610

Zhang G, Silvers L G, Zhao M and Knutson T R 2021 Idealized Aquaplanet Simulations of Tropical Cyclone Activity: Significance of Temperature Gradients, Hadley Circulation, and Zonal Asymmetry *Journal of the Atmospheric Sciences* **78** 877–902

Supplementary Information for

**Characteristics and Predictive Modeling of Short-term Impacts of Hurricanes on the US Employment**

[1] Department of Climate, Meteorology, and Atmospheric Sciences, University of Illinois at Urbana-Champaign, 1301 W Green Street, Urbana, IL 61801, United States

[2] Nanyang Business School, Nanyang Technological University, 91 Nanyang Avenue, Singapore 639956

**1. Data Pre-processing**

For the employment analysis, this study uses the raw data from the BLS without seasonal adjustments. The rationale is that some employment seasonality in cyclone-prone regions can arise from the seasonality of weather (including hurricane activity), so the seasonal adjustments may unnecessarily complicate the interpretation of results. Moreover, our analysis emphasizes the employment shocks immediately after cyclone hits (e.g., 1 month), so the changes attributable to employment seasonality are relatively small. We have also explored applying seasonal adjustments of X-13ARIMA-SEATS from the US Census Bureau. The results from this sensitivity test will be briefly discussed in Section 4.2.

The employment basis and absolute changes range several orders of magnitude for geographic entities, so we use the relative changes to facilitate the comparison of employment changes across regions. The reference basis is the local employment value in the month that immediately precede storm impacts. Large relative changes (e.g., >200%) occasionally happen in small employment markets. But since these cases are extremely rare and not representative of storm impacts, we drop the entries in regions where the goods-producing sector or the service-providing sector employs fewer than 100 individuals. After those steps, we applied a logarithmic scaling when visualizing and analyzing the employment basis. The pre-processing of the wage data follows an identical procedure, except for a linear interpolation of quarterly data to a monthly frequency.

The pre-processing of hurricane data selects the over-land data points and adds the precipitation information. We use a 1:5,000,000 map from the U.S. Census Bureau to identify the storm track data points that are over the US land territories. A caveat of excluding offshore storms is missing the data points where the center of a storm is not over land, but the storm is close enough to affect coastal regions (e.g., Hurricane Harvey 2017 in Section 4). Nonetheless, such cases are overall rare and thus not treated specially.

For the original IBTrACS storm variables, we follow the North American convention and refrain from converting the imperial units (e.g., 1 kt = 0.514444 m s$^{-1}$). To approximate the storm precipitation, we search the CPC precipitation data for the value of the nearest grid point. As a crude approximation, we consider the 3-day total precipitation around the storm passing time and attribute the precipitation to the storm of interest. The choice of a 3-day aggregation window is to account for the spatial-temporal heterogeneity of precipitation around moving storms.

## 2. Method Weakness with Near-shore Storms

A careful examination of the data points in Figure 3 reveals a weakness in the preprocessing of nearshore data. The most extreme example is related to Hurricane Harvey (2017), which stalled near the Texas coastline and caused extreme precipitation (>500 mm) around Houston. Some of the most extreme precipitation observations (>500 mm) were missed by our method (c.f., precipitation values in Fig. 3b) as a substantial amount of precipitation occurred when the storm center lingered over the ocean. Consequently, the over-land check (Section 2.2.1) rejected the related storm data points. However, storms like Hurricane Harvey (2017) are rare in the examined samples, so we do not expect this weakness to undermine the overall findings to be presented in this section.

## 3. Correlation Analysis and Confounding Variables

The storm-employment relationship involves many interconnected factors, so it is likely helpful to orient readers with diverse knowledge backgrounds with a summary of statistical information. The information also helps identify cofounding variables and inform the later analyses. The group of variables is relatively small. We considered incorporating other socioeconomic data (e.g., average income, age, and poverty rate) into the analyses. However, the data frequency, time coverage, and region coding of the data are not well aligned with the employment data. To keep the analysis straightforward, we chose to focus on the variables from the BLS and physical datasets.

For the selected group of variables, we present their histograms of raw data (Supplementary Figure 1) and discuss their relationships using correlations calculated with the standardized values (Supplementary Figure 2). This enables the use of correlation analysis between variables with widely different scales. For example, the wind and pressure metrics show significant correlations with the precipitation accumulation (Supplementary Figure 2a), suggesting a tendency of hazard compounding of intense storms. The correlations between storm features and employment changes are much weaker than among some physical variables, suggesting that storm impacts on employment are overall moderate. Nonetheless, some correlations are statistically significant. For example, the correlations indicate more intense, wetter, and slow-moving storms tend to cause employment losses in service-providing industries. The correlations also suggest greater employment losses in service-providing and goods-producing industries occur with late-season storms (Supplementary Figure 2a).

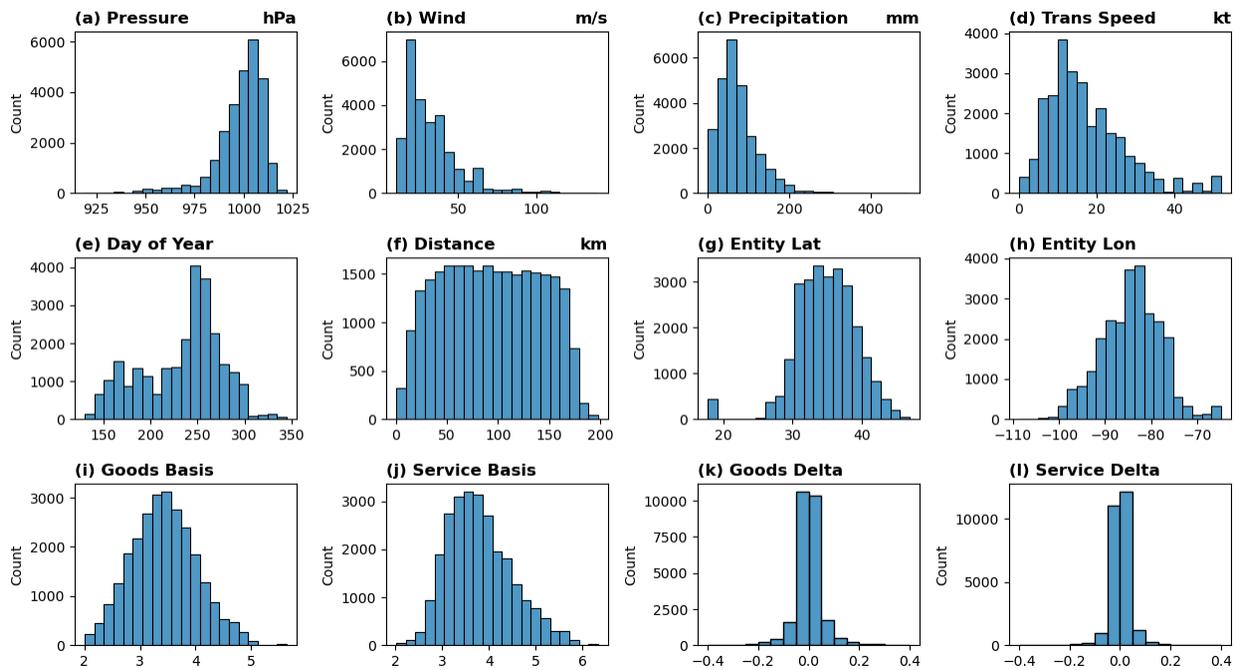

**Supplementary Figure 1 Histograms of storm characteristics and employment data at the county level. The total sample size is approximately 26,200. In the last row, "basis" represents the employment number in the previous month of storm impacts, and "delta" represents the Month-1 fractional changes in employment numbers. The basis employment numbers are scaled with log10.**

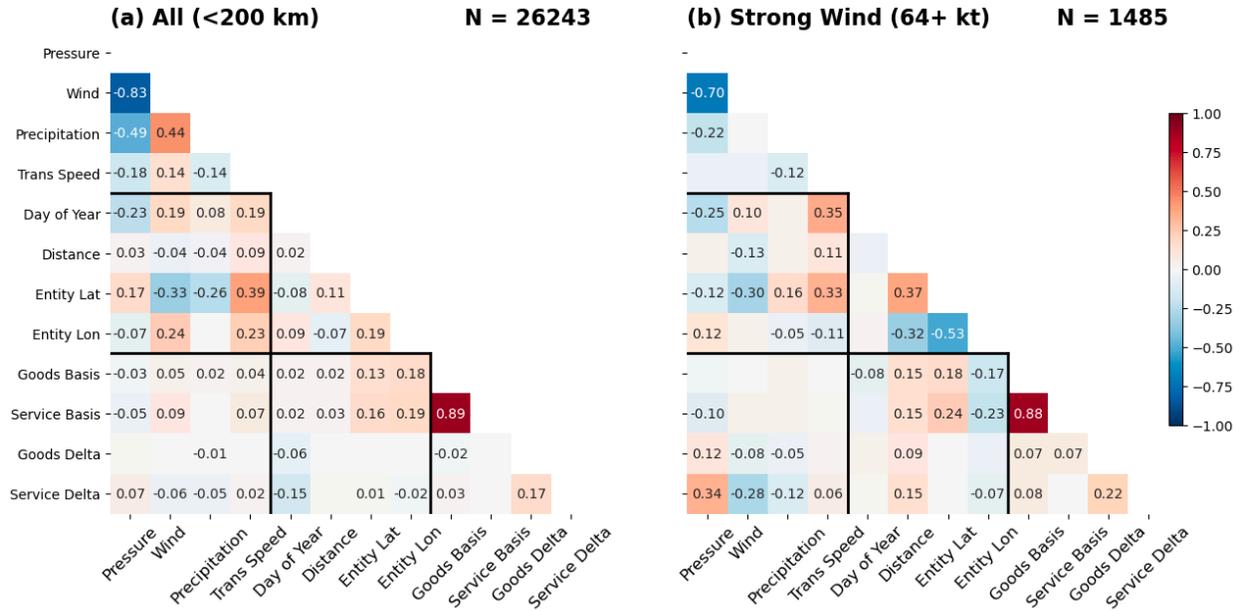

**Supplementary Figure 2 Correlation matrices of storm characteristics, spatial-temporal information, and employment basis and changes. (a) Data points with all the storm cases; (b) Data points with all the hurricane cases (>64 kt). The values in the cells denote correlation coefficients and the colors of the cells denote significance at the 99%-level. Cells with numeric values displayed denote statistically significant correlations.**

The storm-employment correlations are sensitive to the choice of storm samples. When conditioned on intense storms (Supplementary Figure 2b), the correlations between the impact time and employment changes are no more statistically significant, possibly because a conditional check of storm intensity also serves as one of impact time due to the intensity-time correlation. Interestingly, the correlation between employment changes and storm-county distance strengthens. These strong correlations are consistent with the expectation that hazards and damages near the storm center are more severe. The correlation results are qualitatively similar between hurricanes and major hurricanes (>96 kt; not shown).

## 4. Training and Cross-validation of Predictive Models

As the collinearity of variables affects the MLR model and complicates the interpretation of results, we train the predictive models with a reduced set of input variables in Supplementary

Figure 2. Specifically, we drop the minimum pressure of storms and use the wind as the intensity metric exclusively. Additionally, when training models for predicting employment changes in one group of industries (e.g., servicing-providing), the inputs include the employment basis in the same group but not the other group of industries (e.g., goods-producing). Using this reduced subset of variables help remove variables that have strong collinearity, which may contaminate the estimates of model coefficients.

All the models are trained with five-fold cross-validation to mitigate potential overfitting problems. The cross-validation preserves the temporal order as the random sampling inflates skill scores by splitting related data points (e.g., the same storm and adjacent entities) into training and testing groups. This would severely violate the independence assumption of training and testing data. With our cross-validation strategy, the same issue still occurs but affects fewer than 1% of data points in each test. The overall impacts are deemed low and will unlikely undermine our conclusions. Finally, the spread in the coefficient of determination (R2) increases as the sample size decreases (Table 2). In small-sample studies, the overfitting issue could undermine the models training.

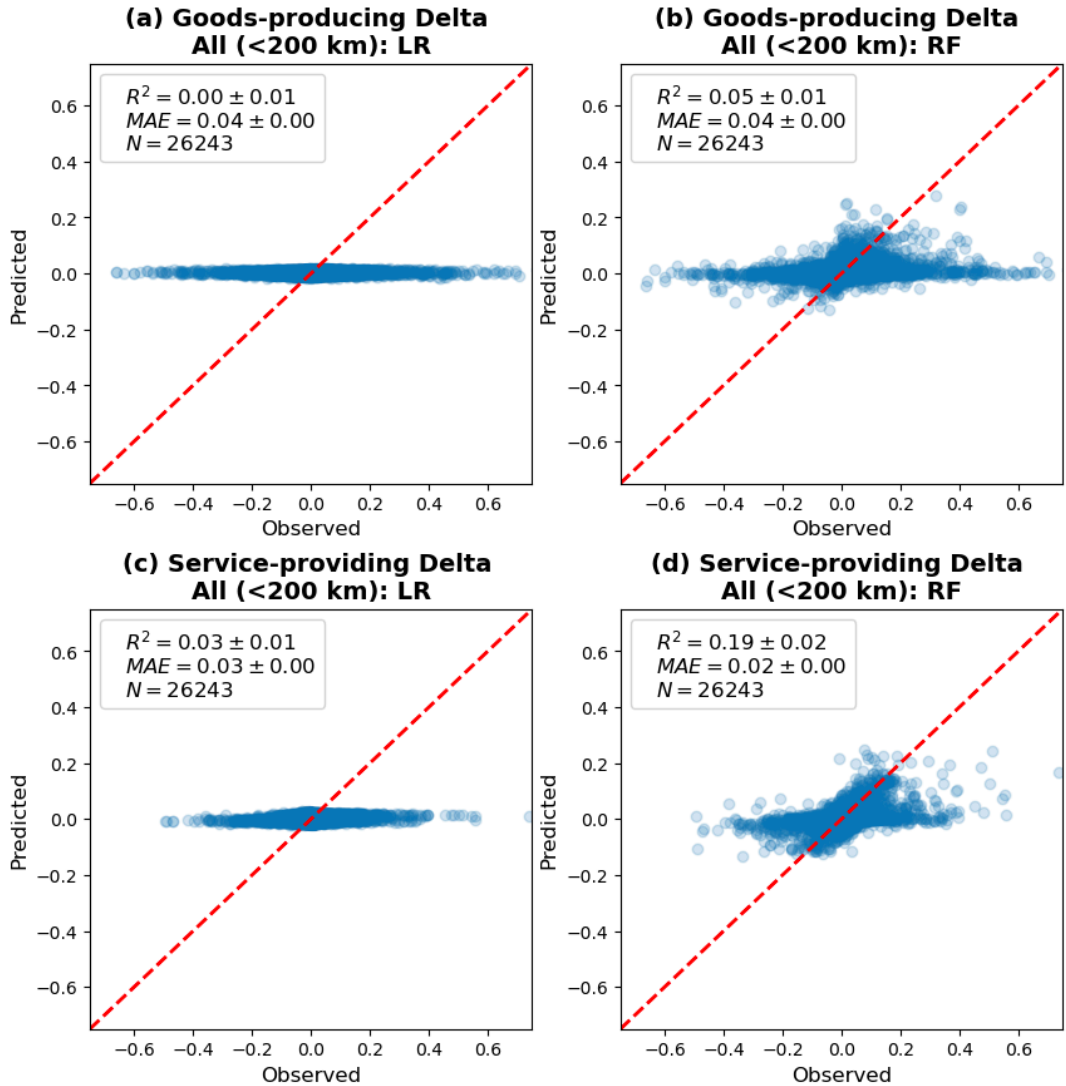

**Supplementary Figure 3 Comparison of the observed and predicted employment changes in county-level employment. (a) Goods-producing sector prediction based on the MLR model; (b) Same as (a) but based on the RF model. (c-d) Same as (a-b) but for the service-providing sector. The blue dots show individual data points. The red line shows the reference line where the prediction matches perfectly with the observation. The coefficient of determination, mean average error, and sample number are denoted in boxes in each panel. The values are shown in the format of mean ± standard deviation based on the cross-validation results.**